\documentclass{vldb}
\usepackage{booktabs} 
\makeatletter
\newif\if@restonecol
\makeatother

\usepackage[linesnumbered,ruled,vlined]{algorithm2e}
\usepackage{algpseudocode}
\usepackage{amsmath}
\usepackage{verbatim}
\usepackage{graphicx}
\usepackage{subfigure}
\usepackage{amsfonts}
\usepackage{bm}
\newcommand{\newblock}{}
\usepackage{url}
\usepackage{listings}
\usepackage{xcolor}
\usepackage{lstautogobble}
\usepackage{balance}
\vldbTitle{DRONE: a Distributed Subgraph-Centric Framework for Processing Large Scale Power-law Graphs}
\vldbAuthors{Xiaolewen, Shuai Zhang and Haihang You}
\vldbDOI{https://doi.org/10.14778/xxxxxxx.xxxxxxx}
\vldbVolume{12}
\vldbNumber{xxx}
\vldbYear{2018}

\begin{document}
\title{DRONE: a Distributed Subgraph-Centric Framework for Processing Large Scale Power-law Graphs}



%
%
%
%
\numberofauthors{3} 
\author{
%
%
\alignauthor Xiaole Wen\\
       \affaddr{SKL Computer Architecture}\\
       \affaddr{Institute of Computing Technology, Chinese Academy of Sciences}\\
       \email{wenxiaole@ict.ac.cn}
\alignauthor Shuai Zhang\\
       \affaddr{SKL Computer Architecture}\\
       \affaddr{Institute of Computing Technology, Chinese Academy of Sciences}\\
       \email{zhuangshuai-ams@ict.ac.cn}
\alignauthor Haihang You{\titlenote{Corresponding author}}\\
       \affaddr{SKL Computer Architecture}\\
       \affaddr{Institute of Computing Technology, Chinese Academy of Sciences}\\
       \email{youhaihang@ict.ac.cn}
}
\maketitle
\begin{abstract}
Nowadays, in the big data era, social networks, graph data-\\bases, knowledge graphs, electronic commerce etc. demand efficient and scalable capability to process an ever increasing volume of graph-structured data.
To meet the challenge, two mainstream distributed programming models, \emph{vertex-centric} (\textbf{VC}) and \emph{subgraph-centric} (\textbf{SC}) were proposed. 
Compared to the \textbf{VC} model, the \textbf{SC} model converges faster with less communication overhead on well-partitioned graphs, and is easy to program due to the "think like a graph" philosophy. The edge-cut method is considered as a natural choice of \emph{subgraph-centric} model for graph partitioning, and has been adopted by Giraph++, Blogel and GRAPE. 
However, the edge-cut method causes significant performance bottleneck for processing large scale power-law graphs. 
Thus, the \textbf{SC} model is less competitive in practice. 
In this paper, we present an innovative distributed graph computing framework, DRONE(Distributed gRaph cOmputiNg Engine). It combines the \emph{subgraph-centric} model and the \emph{vertex-cut} graph partitioning strategy. Experiments show that DRONE outperforms the state-of-art distributed graph computing engines on real-world graphs and synthetic power-law graphs. DRONE is capable of scaling up to process one-trillion-edge synthetic power-law graphs, which is orders of magnitude larger than previously reported by existing SC-based frameworks. 
\end{abstract}

\section{Introduction}
Graph computation becomes vital to a wide range of data analytics such as link prediction and graph pattern matching. Mining "insights" from large scale graphs is a challenging and engaging research. It is a difficult mission to analyze large graphs in parallel on hundreds or even thousands of compute nodes.  Thus, distributed graph processing is increasingly attracting attention from researchers in both academia and industry. To carry out graph computing in a parallel/distributed fashion, high-level programming models such as \emph{vertex-centric} (\textbf{VC}) and \emph{subgraph-centric} (\textbf{SC}) have been proposed. In recent years, respective distributed graph computing systems have been developed including Pregel~\cite{malewicz2010pregel}, Graphlab~\cite{low2012distributed}, Giraph++~\cite{tian2013think}, and etc.

\textbf{VC}, \emph{vertex-centric}~\cite{malewicz2010pregel} tagged as "think like a vertex", is an engineering approach by constructing entire graph computation from programs running on each vertex and its one-hop neighbours connected by inward edges. Although \textbf{VC} has been proved to be general enough to express a broad set of graph algorithms, it does not always perform efficiently. User-defined computing logic is restricted by a single vertex view. The information in the graph flows from one vertex to its one-hop neighbours per iteration. It leads to a large amount of communications and slow convergence. 
Under the \textbf{VC} model, it is mandatory for users to modify the original sequential algorithms to comply with "think like a vertex" philosophy. Furthermore, \textbf{VC} hides the graph partition structure from users, thus forbids the graph-level optimization such as indexing and compression. 
 
On the contrary, \textbf{SC}, \emph{subgraph-centric} (i.e. \emph{block-centric}, \emph{partition-centric})~\cite{fan2017parallelizing}~\cite{tian2013think}~\cite{yan2014blogel}~\cite{simmhan2014goffish}, 
is an elegant and natural choice of programming model for parallel graph computation. Other than extra functionality to deal with information exchange between subgraphs, original sequential code remains without much modification on each subgraph. Such model is tagged as "\emph{think like a graph}".
In addition, since subgraphs are more "coarsely grained" than a single vertex, they retain local connected components in original graph. Hence, \textbf{SC} inherently converges faster and generates less communication traffic. 

However, the performance of \textbf{SC} is highly dependent on sophisticated pre-processing including distributed graph partition and placement. The quality of partitioning plays a central role in minimizing communication and ensuring workload balance. Moreover, such an optimization problem is \emph{NP-hard}. 
Furthermore, graphs generated by modern applications such as World Wide Web and social networks tend to have \emph{highly skewed degree distributions}, which are  categorized as \emph{power-law} graphs. In fact, large scale power-law graphs processing is rarely explored by existing \textbf{SC} frameworks such as GRAPE ~\cite{fan2017parallelizing}, Giraph++~\cite{tian2013think}, GoFFfish~\cite{simmhan2014goffish}. 

To the best of our knowledge, most frameworks employ an \emph{edge-cut} partitioner that assigns vertices to partitions and boundary edges span among partitions(see Section ~\ref{sec:back}). 
For instance, GRAPE ~\cite{fan2017parallelizing}, Graphlab ~\cite{low2012distributed}, and Giraph++ ~\cite{tian2013think} utilize PARMETIS~\cite{karypis1997parmetis} to execute edge-cut partitioning. 
Giraph++ ~\cite{tian2013think} adopts and extends PARMETIS. Blogel ~\cite{yan2014blogel} proposes an edge-cut partitioner that assigns vertices to random \emph{"cell vertex"} by performing multi-source BFS. However, edge-cut partitioner mentioned above all perform poorly on large-scale power-law graphs.
Although Giraph++, a \emph{subgraph-centric} framework, allows users to resort to hash (random) vertex placement for easy to implement and fast graph partitioning, unfortunately, it cuts the most of edges and results in heavy communication and storage overhead. As a result, there will be little to no performance benefit compared to the \emph{vertex-centric} model. 
 Recently, Gonzalez et al demonstrated that vertex-cut partitioner performs well on many large scale real-world graphs with the \emph{GAS} decomposition (Powergraph~\cite{gonzalez2012powergraph}). However, the GAS decomposition is under \emph{vertex-centric} model, and prohibits direct interaction between vertices that are not adjacent in the graph.
 
Given these developments, we believe that there is an opportunity to unify the \emph{subgraph-centric} model and  \emph{vertex-cut} partitioning and build a single system to address the performance issues in large scale real-world graphs processing.   
In this paper, we analyze the limitations of existing subgraph-centric distributed graph computing systems for processing large-scale power-law graphs. We introduce the SVHM abstraction which combines \emph{subgraph-centric} and \emph{vertex-cut} heterogeneous processing scheme. The SVHM abstraction inherits the performance advantages and "Think like a graph" philosophy of the subgraph-centric programing model, and exploits the efficiency and effectiveness of the vertex-cut partitioning method to distribute power-law graphs. We implement SVHM abstraction in a distributed graph processing framework, called DRONE (Distributed gRaph cOmputiNg Engine). We describe the design of our distributed implementation of SVHM such as adaptive computing and subgraph-level communication, and provide a user-friendly programming interface, illustrate its flexibility by presenting the implementation of several classic algorithms with DRONE. We evaluate the performance of DRONE in terms of communication, numbers of iterations, workload imbalance, and running time on popular benchmarks, compare with the state-of-art frameworks.  DRONE's scalability is assessed by processing large-scale synthetic power-law graphs with hundreds billion of edges. 

The rest of the paper is organized as follows. Section~\ref{sec:back} reviews related works. Section~\ref{sec:challenge} is an analysis of the challenges of power-law graphs in existing graph parallel frameworks. Section~\ref{sec:svhm} describes the details of DRONE's SVHM model. Section~\ref{sec:implementation} gives an overview of the DRONE implementation and programming interfaces. Section~\ref{sec:distributedGraphPlacement} discusses the distributed graph placement of DRONE. Section~\ref{sec:applications} exploits DRONE in various graph applications. Section~\ref{sec:experiments} provides a detailed empirical study. In Section~\ref{sec:conclusion}, we conclude our work and describe potential future work.

\section{Background and Related Works}\label{sec:back}
In this section, we first define some notation and have a brief discussion on the general execution procedures of parallel graph computing systems. Later, we review several contemporary parallel graph computing systems with the emphasis on the programming models they expose. The characteristics of these systems are listed in Table~\ref{tab:comp}.
\begin{table*}[htbp]
\begin{tabular}{@{}lllll@{}}\toprule
Systems & Programming Model & Graph partitioning methodology & Execution Model & Communication Paradigm \\ \midrule
Pregel & Vertex-Centric & edge-cut & Synchronous & Message-passing\\ 
GraphLab & Vertex-Centric & edge-cut & Asynchronous & Shared Memory\\
Powergraph & Vertex-Centric & vertex-cut & Asynchronous, Synchronous & Shared Memory\\
Giraph++ & Subgraph-Centric & edge-cut & Synchronous & Message-passing\\
Blogel & Subgraph-Centric & edge-cut & Synchronous & Message-passing\\
GoFFfish & Subgraph-Centric & edge-cut & Synchronous & Message-passing\\
GRAPE & Subgraph-Centric & edge-cut & Synchronous & Message-passing
\\ 
\bottomrule
\end{tabular}
\caption{Distributed graph processing frameworks comparison}
\label{tab:comp}
\end{table*}

\textbf{Notations}
Given a directed graph \begin{math} G(V,E,D) \end{math}, \begin{math} V \end{math} denotes the set of vertices and \begin{math} {E} \end{math} denotes the set of edges. An edge is represented as a pair of vertices \begin {math} (u, v) \end{math} where \begin{math} u \end{math} is the source vertex and  \begin{math} v \end{math} is the target vertex. Vertices and edges might have associated value (metadata) and local data structure of arbitrary type. For \begin{math} v \in V \end{math}, \begin{math} D_{v} \end{math} refers to \begin{math} v's \end{math} associated value and \begin{math} D(u, v) \end{math} refers to the value associated with the edge \begin{math} (u, v)\end{math}.  We define the set of one-hop outward-neighbour vertices of \begin{math} v \end{math} as \begin{math} N_{v}^{out} = \{u | u \in V \land (v, u) \in E\} \end{math} and the set of one-hop inward-neighbour vertices of \begin{math} v \end{math} as \begin{math} N_{v}^{in} = \{u | u \in V \land (u, v) \in E\} \end{math}. For the undirected graph, we replace each undirected edge by two edges with opposite directions.  

\textbf{Graph Partitioning}
\begin{figure}[htbp]
\centering
\subfigure[edge-cut]{
\begin{minipage}[b]{0.8\linewidth}
\includegraphics[width=1\textwidth]{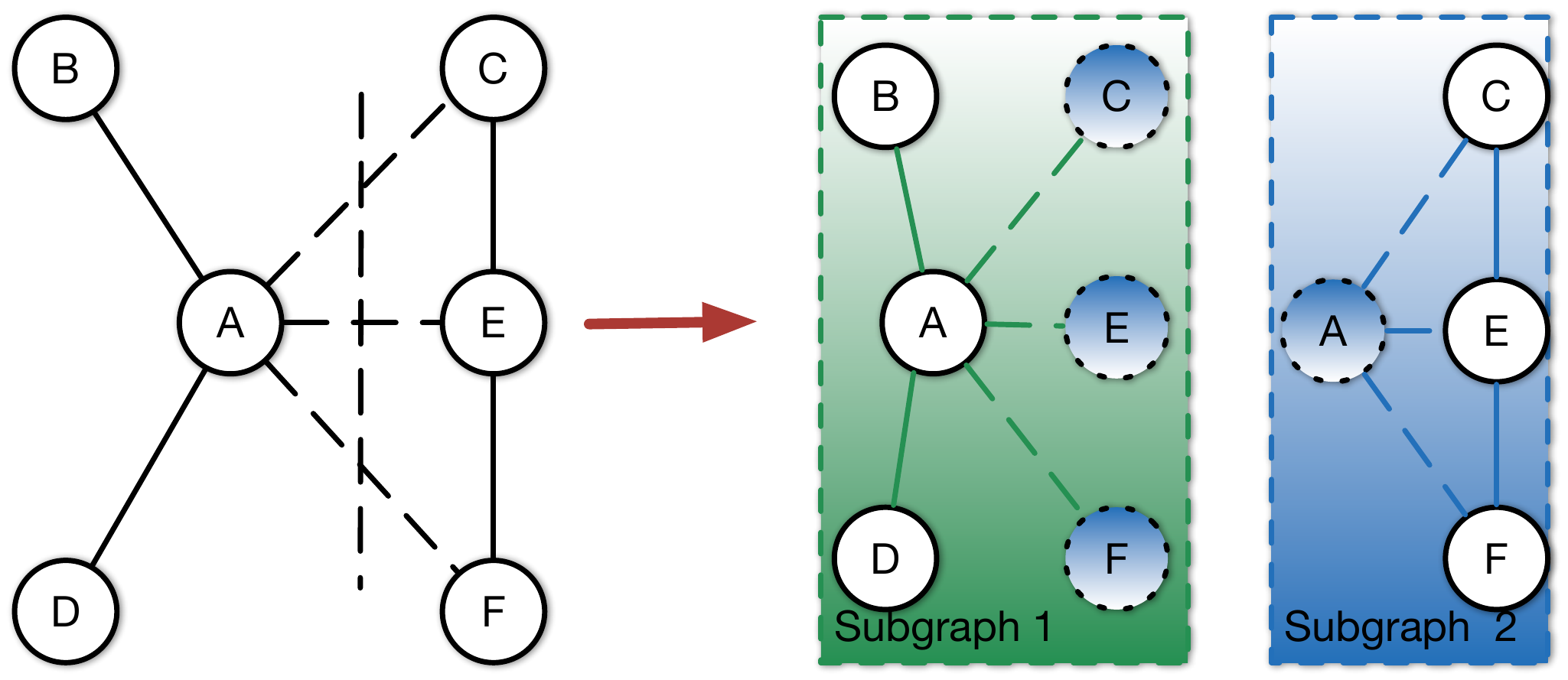}\label{fig.par.a}
\end{minipage}
}
\vfill
\subfigure[vertex-cut]{
\begin{minipage}[b]{0.8\linewidth}
\includegraphics[width=1\textwidth]{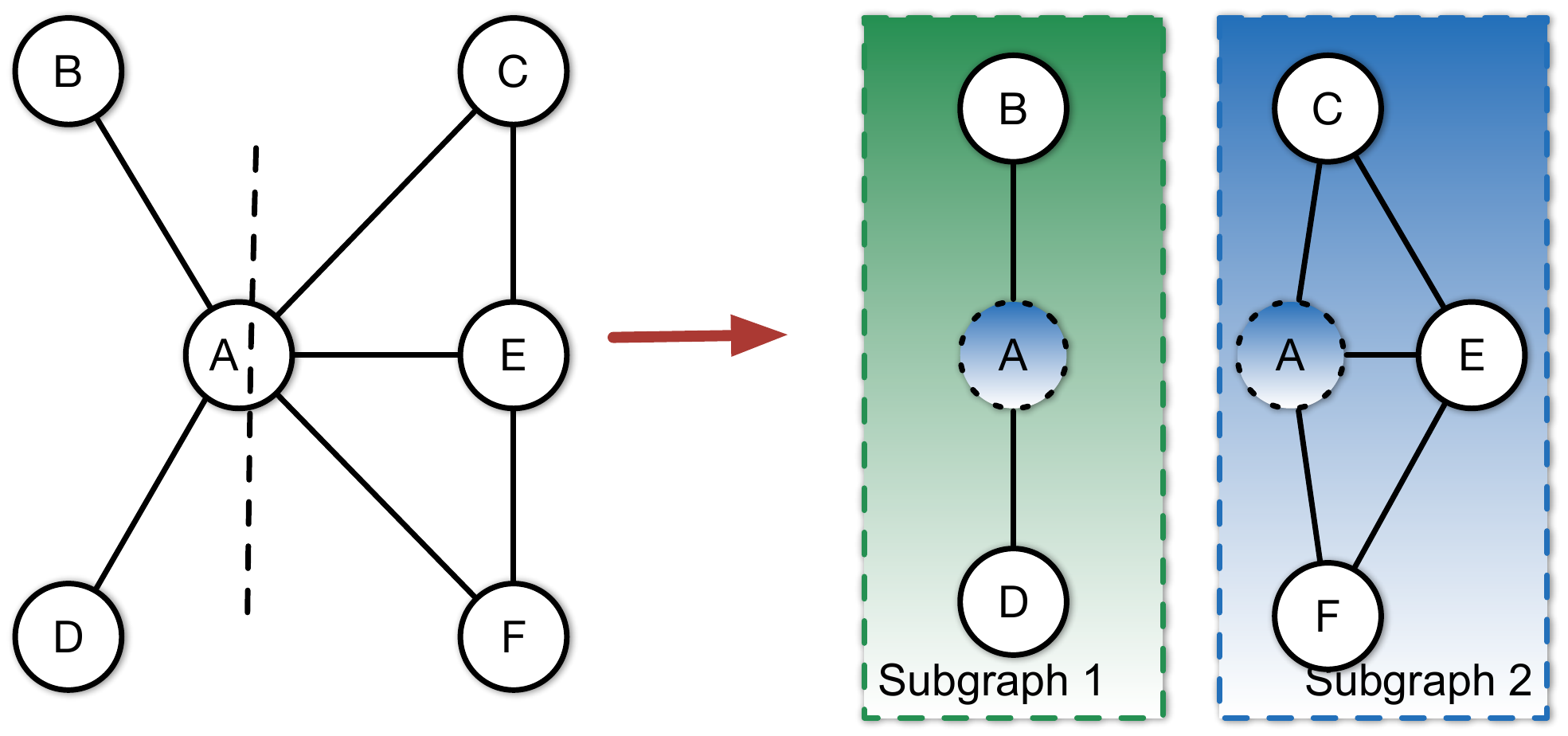}\label{fig.par.b}
\end{minipage}
}
\caption{Two strategies for graph partitioning: \ref{fig.par.a} edge-cut and \ref{fig.par.b} vertex-cut}
\label{fig:partition}
\end{figure}
Graph partitioning(GP) seriously affects the performance of distributed graph computing system in terms of workload balance and communication cost. In addition, the overhead of GP often time is not negligible compared to the total execution time. Existing GP methods can be classified in two main categories: \emph{edge-cut} (a.k.a, \emph{vertex-partitioning}), a classic way, attempts to evenly assign vertices to partitions by cutting the edges, and minimizes the number of cut edges; by contrast, \emph{vertex-cut} (a.k.a, \emph{edge-partitioning}), tries to evenly assign the edges to partitions by cutting the vertices, and minimizes the number of cut vertices. Figure~\ref{fig:partition} shows the difference between the vertex-cut and edge-cut methods. In Figure~\ref{fig.par.a}, the graph is split into two partitions: the vertex sets $\{A,B,D\}$ and $\{C,E,F\}$, where the edges$(A,C)$ ,$(A,E)$,$(A,F)$ are cut. For the edge-cut implementation, each process that handles a partition needs to maintain ghost replications of vertices and edges. In ~\ref{fig.par.b}, the vertex $A$ is cut, and the graph is divided into two partitions which include edge sets $\{(A,B), (A,D)\}$ and $\{(A,F), (A,C), (C,E), (E,F)\}$, respectively. For the vertex-cut implementation, each process that handles a partition keeps a mirror replication of the cut vertex. The ghosts and mirrors are shown in shaded vertices and dotted lines in Figure ~\ref{fig:partition}.  
To our knowledge, the majority of distributed graph computing systems (e.g. ~\cite{malewicz2010pregel}, ~\cite{low2012distributed}, ~\cite{tian2013think}, ~\cite{yan2014blogel}) use edge-cut (a.k.a, \emph{vertex-partitioning}) for GP. However, \emph{vertex-cut} (a.k.a, \emph{edge-partitioning}) has been proposed and advocated in~\cite{gonzalez2012powergraph} as a better approach to process graphs with power-law degree distributions. See Table ~\ref{tab:comp} for a comparison.

\textbf{Pregel/Giraph}
Google proposed a \emph{vertex-centric}(\textbf{VC}) model in Pregel~\cite{malewicz2010pregel}. The kernel of \textbf{VC} is an abstraction that a program \begin{math} Q(v) \end{math} is executed on each vertex $v$ in parallel. Vertices are distributed to different partitions by hashing the vertex ID, and each vertex has all the local data it needs to perform programs: vertex ID, value, a set of outward edges and associated edge value. A complete computing job in \textbf{VC} consists of a sequence of supersteps, which are separated by global synchronization barriers.  At the beginning of a superstep \begin{math} i \end{math},  each vertex gets incoming messages sent by other vertices in superstep \begin{math} i - 1\end{math}, then update the current value of vertex and edges, and send the messages to other vertices for next superstep \begin{math} i + 1\end{math}.  A vertex carries two states: \emph{active} and \emph{inactive}. In the beginning, all vertices are active. A vertex can voluntarily deactivate itself by calling \emph{voteToHalt} or be passively activated by some incoming messages from other vertices. The program terminates when vertices are inactive and there are no pending messages for the next superstep. Apache Giraph~\cite{Giraph} could be considered as the open source version of Pregel, which is also based on the \textbf{VC} model.

\textbf{Graphlab/Powergraph Variant}
Graphlab~\cite{low2012distributed} also carries out the \emph{vertex-centric} (\textbf{VC}) model, which follows the \emph{"Think like a vertex"} philosophy. But unlike the synchronous model and message passing paradigm of Pregel, Graphlab adopts an \emph{asynchronous distributed shared-memory} mechanism, in which the vertex program directly accesses the information in the scope of current vertex, edges and adjacent vertices. Note that shared-memory in this paper is inline with the literature~\cite{DBLP:journals/tkde/KalavriVH18} that is referred as direct data access across vertex-program, message passing is handled implicitly. It is unlike the term used in computer architecture. The workers of Graphlab fetch vertex from the global scheduler in parallel, and add their neighbouring vertices to global scheduler for future execution if necessary. Graphlab ensures serializability by different consistency models, which prevent neighbouring vertex programs from running simultaneously. The asynchronous execution allows the faster worker to move ahead and avoids \emph{straggler effect} but incurs extra locking overhead. The graph structure in Graphlab has to be static, so that no runtime mutations are allowed. 

Powergraph~\cite{gonzalez2012powergraph}, an evolutionary version of Graphlab, proposed a Gather-Sum-Apply-Scatter programming abstraction(GAS).
It tries to address performance issues that arise when using \textbf{VC} model on power-law graphs. Powergraph partitions the graph by vertex-cut instead of edge-cut, so that edges of a high-degree vertex are handled by multiple workers. Accordingly, parallelizing the vertex-centric GAS program over its adjacent edges. PowerGraph programs could be executed both synchronously and asynchronously. 

\textbf{Giraph++}
Giraph++~\cite{tian2013think} proposed the \emph{subgraph-centric} (\textbf{SC}) programming paradigm. The main idea behind the \textbf{SC} model relies on the perception that each partition is a proper subgraph of the input graph, instead of a collection of unassociated vertices. While in the \textbf{vertex-centric} model a vertex is restricted to accessing information from immediate neighbours, the \textbf{subgraph-centric} model allow information to propagate freely between all vertices within a partition. This property of the \textbf{SC} model leads to significant communication improvement and faster convergence.

Giraph++ adopts an \emph{edge-cut} partitioner. Inside a partition, vertices can be \emph{internal} or \emph{boundary}. Internal vertices are associated with their value, neighbouring edges, and incoming messages. Boundary vertices only have a local copy of their associated value, the primary value resides in the partition where the vertex is internal. The messages exchanged between partitions are only sent from boundary vertices to their primary copy. In this way, information exchanged between internal vertices is cheap and immediate. Giraph++ extends PARMETIS for graph partitioning, which is expensive and performs poorly for large-scale Power-law graphs.

\textbf{Blogel/GRAPE Varient}
The \emph{subgraph-centric} (\textbf{SC}) model has been quickly adopted and further optimized in several succeeding works such as Blogel~\cite{yan2014blogel} and GRAPE~\cite{fan2017parallelizing}. Blogel allows a subgraph to define and manage its own state, even subgraph-level communication. 
GRAPE's abstraction essentially decomposes parallel \textbf{SC} programs into separate phases: \emph{Partial Evaluation} and \emph{Incremental Evaluation}, which allows users to plug sequential graph algorithms into GRAPE with minor changes.

\section{Challenges of Real-world Graphs}\label{sec:challenge}
 
Large scale real-world graphs impose great challenges to efficient distributed graph processing. One of the most notable properties is the \emph{skewed} power-law degree distribution: most vertices have relatively few neighbours while a few vertices have many neighbours. Skewed degree distribution is common for natural graphs, such as \emph{Social networks} including Twitter follower network and collaboration networks, \emph{Computer networks} including the internet and the web graph of the World Wide Web. The highest-degree nodes in power-law graphs(also known as \emph{scale-free} networks) are often called \emph{"hubs"}. The probability that a randomly sampled vertex from a power-law graph has degree $d$ is given by: 
\begin{equation} \mathbb{P}(degree = d) = \frac{1}{d^\alpha} \end{equation}
where the exponent $\alpha$ is a positive constant. The lower $\alpha$ is, the more skewed a graph will be. 
In many real-world power-law graphs, the value of constant $\alpha$ is between 2 and 3~\cite{faloutsos1999powergraph16}. 

We examined the performance bottleneck of existing parallel graph computing frameworks based on edge-cut partitioning while processing large-scale power-law graphs. Following are our observations:

(1) \textbf{Workload imbalance}: the traditional edge-cut partitioner tries to evenly assigns vertices, neglects the degree of vertices. As a result, for the \emph{vertex-centric} abstraction, since the storage(memory usage), communication, and computation complexity is linear to the degree of central vertex, the running time of a vertex-program can vary widely~\cite{gonzalez2012powergraph}. For the \emph{subgraph-centric} abstraction, the running time of each subgraph varies and is significantly affected by its total number of edges~\cite{tian2013think}. The subgraphs that contain high-degree vertices will create heavy workload and lead to straggler effect. To address such workload imbalance problem caused by evenly assigning vertices, Giraph++ and Blogel would generate a significantly larger number of partitions than the number of workers, each worker will take numbers of partitions to process. Some frameworks utilize Parmetis' mutli-constraint partitioning feature to generate partitions so that the number of edges are balanced. However, both these approache further increase the overhead of partitioning.

(2) \textbf{Partitioning overhead}: Partitioning overhead is another essential issue for large-scale graph processing. Both the \emph{subgraph-centric} and \emph{vertex-centric} based frameworks depend on the graph partitioning to ensure load balance and minimize communication.       
However, most edge-cut partitioning algorithms perform poorly on power-law graphs. For example, the publicly-available software Parmetis, which is adopted by Graphlab and GRAPE, is not able to finish partitioning a power-law graph with one billion edges within six hours on our platform. Giraph++ extends the Parmetis, which is still an expensive method. While the power-law graph is difficult to partition, Giraph++ and Graphlab allow users to resort to random hash partitioning. However, although hash partitioning is easy to implement and fast, it destroys the local connected structure of original graphs. 

(3) \textbf{Communication and storage}: 
For an edge-cut partitioning based graph processing system, assignment of a hub will produce notable overhead of communication and storage: since each inner vertex of a subgraph locally maintains its entire adjacency information, in some cases, including ghost replication of vertices and edges to ensure the update of inner vertices without external memory access. For example, given a toy power-law graph in Figure ~\ref{fig.par.a}, the hub vertex A is assigned to machine 1. We can see that more than half of the edges are cut, and it leads to replication of four vertices and three edges across only two partitions.

\emph{Large diameter(i.e. longest shortest path)} is another property of real-world graphs. It is common for road networks and large Web graphs. Since the information in a graph flows from one vertex to its one-hop neighbour per iteration in \emph{vertex-centric} model, it often leads to slow convergence while processing large diameter graphs with \emph{vertex-centric} based frameworks like Pregel and Graphlab(see Section ~\ref{sec:experiments}).
\section{SVHM Model}\label{sec:svhm}
In this section, we introduce our new graph-parallel model: \emph{subgraph-centric programming and vertex-cut partitioning heterogeneous model (SVHM)}. Briefly, SVHM opens up the entire subgraph of each partition to the user-function, retaining the inherent advantages of the \emph{subgraph-centric} model such as less communication overhead and faster convergency. By integrating the \emph{vertex-cut} partitioning scheme into the model, SVHM is able to process the large-scale \emph{power-law} graphs efficiently. 
\subsection{Preliminaries}\label{svhm:pre}
Give an input directed graph denoted as $G = (V, E)$, an \emph{edge-partitioning} (\emph{vertex-cut}) of $G$ divides all edges into disjoint partitions $E_{1}, E_{2},..., E_{n}$. i.e.  
\begin{equation} 
E = \bigcup_{i=1}^{n} E_i,  \forall i, j : i \neq j \Rightarrow E_i \cap E_j = \emptyset    
\end{equation}
We say that a vertex $v$ belongs to partition $E_i$ if \emph{at least one} of its adjacent edges belongs to $E_i$. Consequently, the $E_i$ is associated with a vertex set $V_i$, composed of the end points of its edges:
\begin{equation}
V_i  =  \{ u | (u, v) \in E_i \lor (v, u) \in E_i) \}
\end{equation}
For the isolated vertices that have no adjacent edges, we just randomly assign these vertices to partitions.
The edges of each partition, together with the associated vertices and assigned isolated vertices, as well as the data in associated with the vertices and edges form the data subgraph $G_i = (V_i, E_i, D_i)$ of $G$.
In the rest of the paper, the terms subgraph and partition are interchangeable. 
Hence, an edge owner's id (aka. a subgraph id) is denoted as $\pi (e) \in \{1,2,...,p\}$, and let $\pi(v) \subseteq \{1,2,...,p\} $ be the set of subgraphs that the vertex replicas reside in.   

Note that while an edge can only belong to one partition, vertices may be replicated among several partitions. The vertices that appear in more than one partitions are called \emph{frontier vertices}, referred to as $F_i$. In contrast, the vertices that appear in just one partition are called \emph{internal vertices} or \emph{no-cut vertices}. For example, in Figure ~\ref{fig.par.b}, A is the frontier vertex, while B and D are internal vertices. For each vertex $v$ with multiple replicas, one of these replicas is designated as the $master$ vertex.
All remaining replicas are treated as $mirror$ vertices, and each maintains a local cached copy of its data.
Consequently, the set of frontier vertices in a subgraph is divided into two subsets: the master vertices set $MA_i$ and the mirror vertices set $MI_i$.
A master maintains a record of the partition ids that its remote mirrors reside. A mirror keeps the partition id of its master. Such a \emph{distributed routing table} mechanism enables \emph{subgraph-level} communication.

Under the \emph{subgraph-centric} model, given $k$ available workers among clusters, we need to decide $n$, the number of partitions of input graph. As we discussed in Section ~\ref{sec:challenge}, Giraph++ and Blogel employ the "one worker to many partitions" (i.e, $n \gg k$) strategy to achieve workload balance when processing Power-law graphs. However, as the number of partitions increases, the average size of the partitions decreases, thus decreasing the chance of a vertex directly accessing its neighbourhood within a single superstep and degrading the performance gain of \emph{subgraph-centric} model.
 For many graph algorithms, the runtime is dominated by the number of edges. Additionally, the memory consumption to store a partition is strictly proportional to the number of its edges ~\cite{guerrieri2014distributed}.  
Unlike the Giraph++ and Blogel, we utilize the \emph{vertex-cut} partitioning in the SVHM model.
The partitioner uniformly assigns edges among workers. Therefore, with the "one worker to one partition" strategy, load balance could be achieved naturally in the SVHM model. 

A subgraph $G_i$  is managed by a process $P_i$. These processes are also called \emph{workers}, and one worker is designated as the \emph{coordinator} to manage the remaining workers. 
Computation proceeds in supersteps which follows the \emph{bulk synchronous parallel model} (\textbf{BSP}). At each superstep, all active subgraphs execute an identical \emph{user-defined-function} (\textbf{UDF}) in parallel. 
At the end of each superstep, the underlying run-time system takes care of the \emph{subgraph boundary synchronization}. Communication takes place at the subgraph-level, and message exchange is implemented in batch fashion. The task terminates when all subgraphs have no pending messages and vote to halt. 
We finally chose BSP for parallel model mainly for its straightforwardness to scale. According to the Facebook experimentation ~\cite{ching2015one}, BSP is easy to identify the system infrastructure and user application bugs, provides the repeatable and reliable results, and simple to take snapshots.

\subsection{User-defined Function}
Computation in the SVHM abstraction is carried out by a \emph{subgraph-centric program} which is a user-defined function(\emph{UDF}) that follows the standards of DRONE's API(see Section ~\ref{sec:api}).
At each superstep, a compute function operates on subgraph $G_i$ with the incoming messages $M_i$. It generates an updated subgraph $G_i^{'}$ as well as a set of $\Delta D_i$ of \emph{frontier vertices} that need to be propagated to other workers, see Equation ~\ref{eq:udf}. To keep the data consistency of the frontier vertices in different partitions, $\Delta D_i$ will be sent off to corresponding replicas of the \emph{frontier vertices}, which is handled by \emph{subgraph boundary synchronization} described later in Section ~\ref{sec:sbs}.  
\begin{equation}\label{eq:udf}
{\bm{Compute}}: f(G_i, M_i) \longrightarrow (G_i^{'}, \Delta D_i)  
\end{equation}
The \emph{SVHM} abstraction allows the \emph{UDF} to have unlimited access to the data of the vertices and edges, and treat the frontier vertices and no-cut vertices equally in computation. Such equal treatment between vertices in a subgraph is the critical feature of \emph{SVHM} that differs from traditional edge-cut based subgraph-centric models. For instance, in Giraph++, the ghost vertices (e.g., C,E,F on machine 1 in Figure ~\ref{fig:partition}) are treated as \emph{"second-class"} in the subgraph. They only keep a temporary local copy of vertex value, which are caches of  the intermediate updates from the internal vertices and need to be propagated to their primary copies (i.e., internal vertices in other subgraphs) through communication. In contrast, the frontier vertices value of the SVHM model are local state in their subgraphs and they are reconciled by the underlying run-time periodically. Users can use the vertices and edges value to accumulate updates for $\Delta D_{i}$(e.g. PageRank in section ~\ref{sec:pr}), or use their vertex values directly (e.g. Algorithm ~\ref{alg:cc}).

\subsection{Subgraph Boundary Synchronization}\label{sec:sbs}
In this section, we introduce the \emph{subgraph boundary synchronization} (SBS) scheme in SVHM. As discussed in Section ~\ref{svhm:pre}, frontier vertices are divided into two categories: master vertices ($MA_i$) and mirror vertices ($MI_i$). In SVHM model, messag exchange between partitions is carried out in two steps: \emph{Aggregate} and \emph{Disseminate}. At the end of each superstep, with user-defined action masters aggregate data from corresponding mirrors. Consequently, masters disseminate updated data to mirrors. Therefore, data coherence is ensured across partitions. More importantly, this boundary synchronization scheme is transparent to end users, which is a distinct feature of SVHM. 

In a nutshell, the SVHM's SBS scheme borrows the idea of parameter server abstraction ~\cite{li2014scaling} and extends this abstraction to support large scale workloads in graph computing. 
The vertex-cut partitioning strategy adopted by SVHM naturally fits in the parameter server model: the status of the master replicas of all frontier vertices can be considered as global states stored in the parameter server. As we mentioned in Section ~\ref{svhm:pre}, the SVHM model randomly designates one of the replicas of each frontier vertex as master vertex, and other replicas as mirrors. Each worker plays dual roles: the \emph{server} and \emph{client}. When global synchronization is triggered, each worker first acts as a client, identifies those pairs which keys in the $MI_i$ from the $\Delta D_{i}$, reconstructs the pairs to a set of messages $M_j$ designated to worker $P_j$ for $j = 1,2,...,n$, the $M_j$ can be represented as a \emph{(key, value)} pairs such that $\{(v.id, v.\delta), v \in MA_j\}$. The worker then sends the $M_j$ to $P_j$. Upon receiving all designated messages by a worker, it acts as a server to assemble the updated data from other partitions into a final result. More specifically, for a master vertex $u$ in subgraph $G_j$, it combines a set of received data $\{\delta_{i}, i = 1,2,...,k\}$ designated to itself with its own $\delta_{j}$ to update vertex value $D_{u}$, applying a user-defined \emph{Aggregate} function for the merge. For example, in the CC algorithm, the \emph{Aggregate} function is \emph{min()}, and \emph{sum()} in the PageRank algorithm. Then each sever \emph{disseminates} the merged data to clients, which concludes subgraph boundary synchronization phase at a superstep.

\begin{figure}[ht]
\centering
\includegraphics[width=0.9\linewidth]{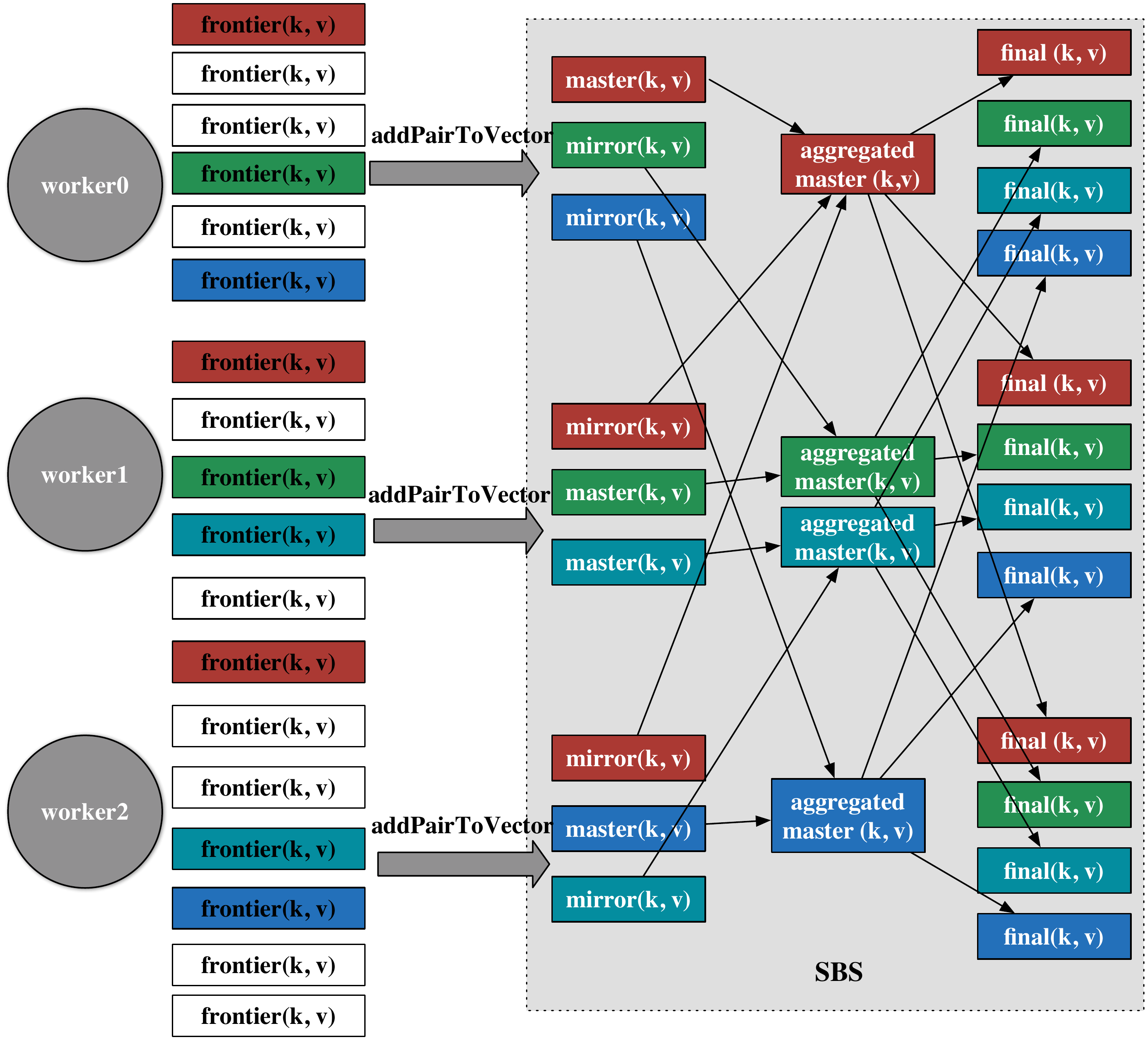}
\caption{Diagram of Subgraph Boundary Synchronization (SBS)}
\label{fig:SBS}
\end{figure}

Figure ~\ref{fig:SBS} is a schematic diagram showing how SBS works. At the end of a superstep, each worker assembles a \emph{(key, value)} vector from frontier vertices with updated data (colored in Figure ~\ref{fig:SBS}). The shaded box shows the SBS procedure that those workers with master vertices will execute \emph{Aggregate} function to consolidate the difference of masters and mirrors, and distribute results to mirrors by \emph{Disseminate} action. Since master vertices are randomly elected, the aggregation workload is evenly distributed across workers.
While Pregel ~\cite{malewicz2010pregel} and Blogel ~\cite{fan2017parallelizing} perform the aggregation on a single designated master, SVHM's SBS alleviates such a scalability issue with evenly distributed masters.

\section{DRONE}\label{sec:implementation}
DRONE, a Distributed gRaph cOmputiNg Engine, is an implementation of the SVHM model developed in the Go programing language. The \emph{Golang} is employed for its high performance networking and multiprocessing, static typing and productivity of building distributed systems. 
\subsection{DRONE Programming API}\label{sec:api}
In this Section we describe the DRONE programming API. Listing ~\ref{list:API} illustrates the core class and a number of major functions of the SVHM abstraction. 

For each subgraph, a look-up table of indexed vertices is constructed during the initialization phase by \emph{Subgraph<>}. \emph{Vertex()} provides a mechanism to query the properties of a vertex including vertex value and its adjacent edges that are exclusive inside the current partition. In addition, we provide a method (\emph{getDegree()}) to query the full degree of a vertex. The function \emph{addPairtoVector()} allows user to build frontier vertices' key-value pairs that need to be propagated to other partitions at a superstep. Moreover, unlike the communication mechanism of subgraph-centric framework Giraph++ ~\cite{tian2013think}, which uses vertex-level message passing and still requires users to explicitly perform sending operations in user-defined computing logic, the message exchange of DRONE is in a batch manner and performed by DRONE runtime seamlessly without user intervention. However, the user needs to construct the function \emph{Compute()} to instantiate the serial graph algorithm on a graph or subgraph. 

 Algorithm ~\ref{alg:cc} elaborates the implementation of the connected components (\textbf{CC}) algorithm with DRONE. Here, the input of CC is an undirected graph in which each vertex takes its unique ID as its label, and when the detection of connected components is done, the label of each vertex is the smallest vertex ID in its corresponding connected component. The sequential CC algorithm is straightforward and well-studied by using either \emph{breadth-first search} or \emph{depth-first search}. Due to the subgraph-centric feature of the SVHM model that user-defined program can random access the entire subgraph, ones can directly reuse the sequential algorithm with DRONE. Specifically, at superstep 0, the sequential connected components algorithm is executed to initiate the local connected components. Then, a \emph{(key, value)} vector is constructed by \emph{addPairtoVector()} if any label change is detected on any frontier vertex. Later on, CC is applied to every vertex of $M$ iteratively at each superstep, where the input $M$ is the updated vertex-label pairs of frontier vertices received from external partitions. Again, label changes of frontier vertices are recorded and propagated by the \emph{SBS} mechanism of the DRONE run-time.
 \lstset{
	frame=single,
	autogobble=true,
	basicstyle=\small\ttfamily,
	}          
\begin{lstlisting}[caption={DRONE programming API}, captionpos=b, label={list:API}]  % Start your code-block

type Subgraph interface {
    GetVertex(id int64) Vertex
    GetVertices() map[int64]Vertex
    GetFrontierVertices() map[int64]Vertex
    GetParents(id int64) map[int64]Vertex
    GetChildren(id int64) map[int64]Vertex
    GetEdge(id1, id2 int64) edgedata
} 
type Vertex interface{
    GetID() int64
    GetData(id int64) vertexData
    GetDegree(id int64) int
}
Compute(g Subgraph, M message) map[key]value
addPairToVector(k key, v value) map[key]value 
getSuperstep() int
voltToHalt() 
\end{lstlisting}
With DRONE, the information exchanged among workers can be represented as a collection of \emph{(key, value)} pairs. As shown in the CC algorithm, the \emph{key} is the vertex id and the \emph{value} is the corresponding vertex label. We expose the \emph{addPairToVector()} interface to users for customization of the transmitting messages according to specific algorithms. Another important role of \emph{addPairToVector()} is that by choosing certain frontier vertices to be reconciled during the SBS, DRONE could reduce communication and support \emph{adaptive computation} to potentially accelerate convergence for iterative algorithms. 
 \begin{algorithm}
  \caption{User-defined function \emph{Compute()} of CC algorithm}
  \label{alg:cc}
  \DontPrintSemicolon
  \KwIn{Subgraph $G$, Key-value pairs $M$}
  \KwOut{Modified subgraph $G^{'}$, Updated key-value pairs $\Delta D$} 
  Vector $vec \gets \emptyset$\;
  \If{getSuperstep() == 0}
   {
  	\textbf{SequentialCC}(root)\;
	\ForEach {v $\in$ getFrontierVertices()} {
		\If{v.labelHasChanged()}{
		 	vec.addPairToVector(v.id, v.label)\;
		}
  	}
  }
  \ElseIf{superstep > 0}{
  	\ForEach {v $\in M$} {
    	 \textbf{SequentialCC}(v)\;
  	}
	\ForEach {v $\in$ getFrontierVertices()} {
		\If {v.labelHasChanged()}{
		 	vec.addPairToVector(v.id, v.label)\;
		}
	}
  }
  \Return vec\;
\end{algorithm}
\subsection{DRONE Execution Workflow}
\begin{figure}[h]
\centering
\includegraphics[width=0.9\linewidth]{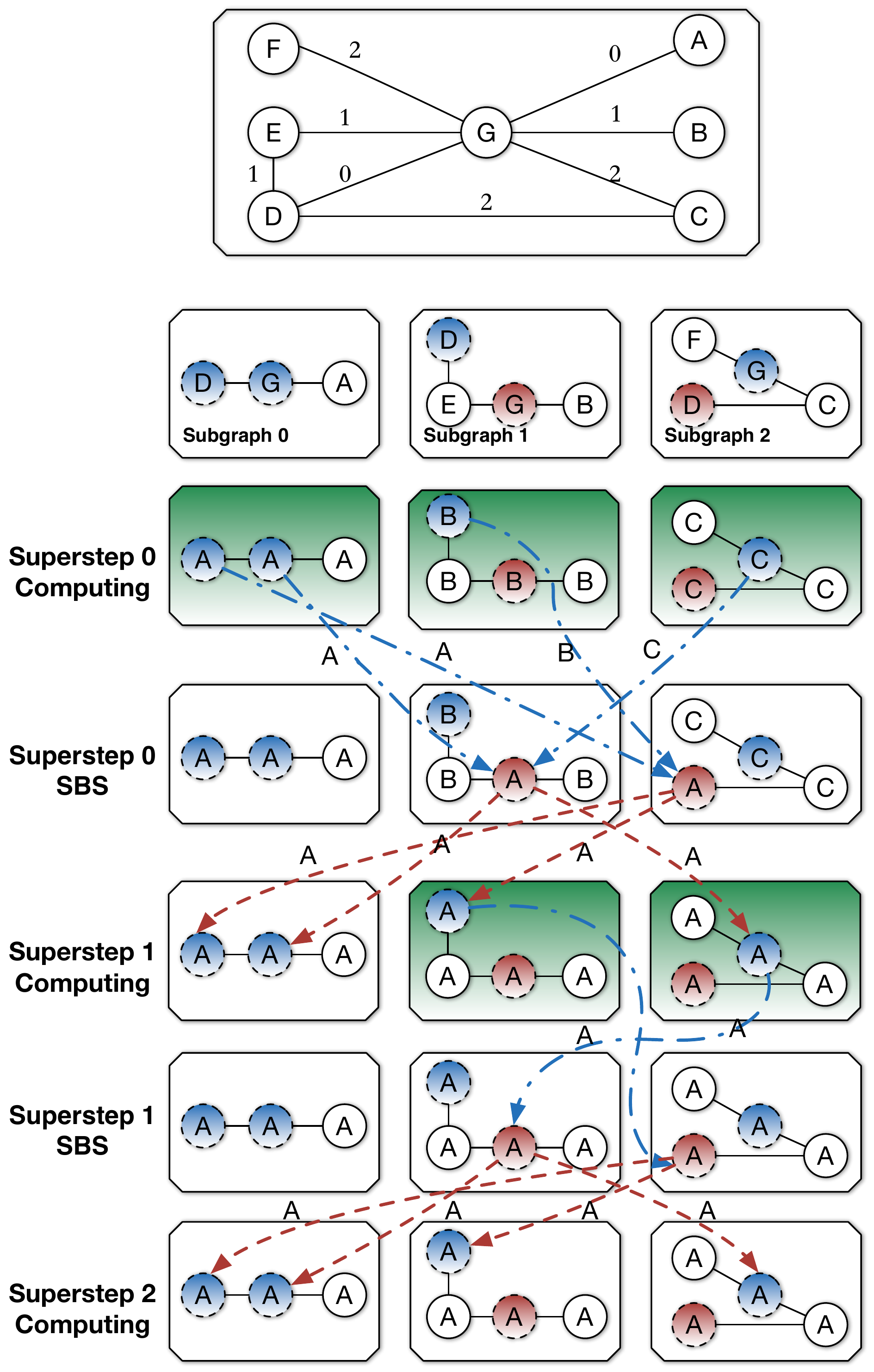}
\caption{DRONE execution workflow of the CC algorithm. Dotted lines are messages. The subgraphs marked in green have messages to be propagated}
\label{fig:cc}
\end{figure}

Figure ~\ref{fig:cc} depicts the DRONE execution workflow of CC algorithm. The input undirected graph contains seven vertices and eight edges, where $\{D,G\}$ are high degree vertices. We employ the vertex-cut algorithm discussed in Section~\ref{sec:distributedGraphPlacement} to divide the input graph into three \emph{subgraphs 0-2} by splitting vertices $\{D,G\}$ and assigning each edge to its owner subgraph (each edge is tagged with its owner subgraph ID in Figure ~\ref{fig:cc}). The frontier vertices are marked with dotted lines, and the \emph{master} and \emph{mirror} vertices are colored in \emph{red} and \emph{blue}, respectively. As we mentioned earlier, the basic idea of programming with the SVHM model is that the sequential CC algorithm can be applied to each subgraph, then merge local \emph{small} components by \emph{subgraph boundary synchronization (SBS)}. As shown in Figure ~\ref{fig:cc}, at the computing phase of superstep 0, each vertex is labelled with the smallest ID in the corresponding connected component within its subgraph. Then the DRONE run-time takes over. The masters aggregate the locally \emph{changed component label} from their mirrors. For the example in Figure ~\ref{fig:cc}, master vertex $G$ with local label $B$ in subgraph 1 receives values from mirrors in subgraph 0 and 2, and their local labels are $A$ and $C$ respectively. The final value of master vertex $G$ is set to be $A$ at the SBS of superstep 0. For the same reason, master vertex $D$ in subgraph 1 is labelled with $A$. Both masters disseminate its updated value to their mirrors.
 In the subsequent supersteps, the CC algorithm propagates the updated label of frontier vertices(including the master and mirror) to their local neighborhood. At the computing phase of superstep 1, the vertices' component label of subgraph 1 and subgraph 2 also changed to A. Similar to Pregel, DRONE has a subgraph-level \emph{voteToHalt()} mechanism: if there is no label change of frontier vertices in the subgraph, the worker will halt. On the other hand, the subgraph will be activated if there are incoming messages. The whole job will terminate if all subgraphs have no pending messages and halt simultaneously.
 
\section{Distributed Graph Placement}\label{sec:distributedGraphPlacement}
The graph partitioning is an essential step in distributed graph processing which greatly affects overall performance. In this section, we discuss the motivation for replacing the edge-cut with vertex-cut in the subgraph-centric programming model, performance metrics for the vertex-cut partitioner, and introduce a scalable \emph{canonical degree-based hashing}  \textbf{(CDBH)} partitioner as the default partitioner of DRONE. 
\subsection{Merits of Vertex-cut Partitioning}
\emph{Vertex-cut} partitioning(also called as edge-paritioning) has been proposed and advocated ~\cite{gonzalez2012powergraph} as a better strategy to divide \emph{Power-law} graphs. For example, Figure ~\ref{fig.par.b} shows a partitioning strategy by cutting the hub vertex A, so that only vertex A with associated data is replicated and needed to be synchronized across the network. Since each edge is stored exactly once, there is no need for data exchange for edges. Compared to the \emph{edge-cut} partitioning mentioned in Section ~\ref{sec:challenge}, \emph{vertex-cut} can significantly reduce communication and storage overhead. The theoretical analysis of the improvements presented in ~\cite{bourse2014balanced} and ~\cite{gonzalez2012powergraph}: for the policy of \emph{randomly} assigning an edge or a vertex to one of subgraphs, the expected communication and storage cost of \emph{vertex-cut} is less than \emph{edge-cut}, the gain of \emph{vertex-cut} has a lower bound and is related to the power-law constant $\alpha$. In addition, Gonzalz et.al~\cite{gonzalez2012powergraph} has proved that for a given good \emph{edge-cut}, it can be easily converted to a better \emph{vertex-cut}. 
\vspace{8pt}

\subsection{Metrics of Vertex-cut Partitioning}
A large number of dedicated \emph{vertex-cut} algorithms have been proposed in recent years. It is a difficult task to choose an appropriate partitioner in practice due to the complexity of evaluating a graph partitioning algorithm. ~\cite{mykhailenko2017distributed} summarizes the existing \emph{vertex-cut} partitioning algorithms and classifies these partitioners. The partitioning quality can be indicated by two categories of metrics: partitioning metrics and execution metrics. 
\paragraph*{Partitioning Metrics}
One category is the statistical data that can be directly computed during the partitioning phase. These metrics basically reveal two aspects: \emph{balance} and \emph{communication}. The balance describes that the size of subgraphs should be approximative. The communication describes that the overlap (e.g., the replicated edges or vertices) across subgraphs should be minimized.The overlap between subgraphs finally contributes to the inter-node communication and storage. We formulated these two metrics for the \emph{vertex-cut} partitioning algorithms as follows: the \emph{Imbalance} is defined as the maximum number of edges in a subgraph divided by the average number of edges across all the subgraphs, i.e. 
\begin{math}
\frac{\max_{i=1,...,n} |E_{i}|}{|E|/n}
\end{math}
 ; the \emph{Replication Factor} is defined as the ratio of the number of vertices in all subgraphs to the total number of vertices in the input graph, i.e.  
\begin{math}
\frac{\sum_{i=1}^{n} |V_{i}|}{|V|}
\end{math}.
\emph{Replication Factor} measures the overhead caused by the fact that some vertices span among subgraphs.
\paragraph*{Execution Metrics}
Since we are focusing on designing a programming model and building a graph processing framework in this paper, we are more concerned about the final effects of the partitioner to the graph processing algorithms (e.g. CC, PageRank). We call the final effects to measure a partitioner are \emph{execution metrics}. The execution metrics mainly include (1) \emph{running time}: it is the time of a graph processing algorithm executed on partitioned graph; (2) \emph{network message}: it is the number of messages (e.g., the number of \emph{(key, value)} pairs in SVHM abstraction), or total bytes transferred by network during the processing; (3) \emph{partitioning time}: the partitioning time measures a partitioner's efficiency. In a production environment, the time consumed during the partitioning phase should be taken into account in the total execution time for a complete processing pipeline. Hence, there is always a trade-off between the efficiency and effectiveness of a partitioner. 
\subsection{Canonical Degree-based Hashing Vertex-cut}\label{cdbh}
We adapted the vertex-cut algorithm proposed in ~\cite{xie2014distributed} into a scalable canonical degree-based hashing scheme and made it as the default partitioner of DRONE.
DRONE supports several built-in vertex-cut partitioning algorithms. The partitioner of DRONE also follows a distributed manner. DRONE supports two input formats: (1) \emph{vertex-oriented}: one data entry consists all data related to a vertex, including outgoing edges; (2) \emph{edge-oriented}: each item is an edge data entry, and vertex data is stored separately. For the second format, DRONE will group edges by their source vertices in a MapReduce job for pre-processing. Then each partitioning worker loads a collection of vertices with associated data from external storage. For instance, worker $i$ holds the vertex $v$ that $hash(v) = i$. Then workers compute the \emph{edge-to-worker} assignment, i.e, each outgoing edge's owner of the vertices it holds. In the original DBH algorithms, the owner of an edge is determined by the hash function value of the associated vertex with smaller degree. Hence, to compute the \emph{edge-to-worker} assignment, each worker needs to retrieve the degree of the adjacent vertices of the vertices it holds, then determining the owner of the corresponding edge. When the \emph{edge-to-worker} assignment is computed, each worker can deduce where its vertices' replications reside by referencing its outgoing edges' owner list. Next, workers exchange the vertices and edges data and location metadata to their owner, and each worker constructs subgraphs by received data as the paradigm described in Section ~\ref{svhm:pre}. Every constructed subgraph will be dumped into a separate file on a distributed file system. Compute workers will load these subgraphs for distributed processing. Another key extension is that we assign an edge in a "canonical" way, meaning that we sort the two associated vertices by their vertex id first and then apply the function $f$. The sorting ensures assigning the opposite directed edges $(u,v)$ and $(v,u)$ to the same subgraph. Note that the CDBH partitioner works very fast for the intrinsic hash, but we also evaluate the execution performance of CDBH in Section ~\ref{sec:partitionercompare}. \vspace{10pt}

\section{Applications}\label{sec:applications}
In this section, we show the implementation of three representative classic graph algorithms in the \textbf{SVHM} model under the DRONE framework.
\subsection{Single Source Shortest Path}
Given a graph $G(V,E)$ and a source vertex $s$, single source shortest path (SSSP) computes the length of shortest path $dist(s, v)$ for all $v \in V$. 
Dijkstra has proposed a sequential algorithm in $O((|V|+|E|)log|V|)$ time complexity for this problem ~\cite{fredman1987fibonacci}.
In Dijkstra's algorithm, a \emph{min-priority queue} is maintained for storing vertices associated with the
latest updated shortest path length from source vertex $s$.

In the \textbf{SVHM} abstraction, due to the \emph{"think like a graph"} philosophy, we keep the sequential Dijkstra's algorithm to compute the $dist(s, v)$ for each $v \in G_i$, and maintain a vector of $(v.ID, dist(s,v))$ where each $v$ is a frontier vertex. If $s \notin G_i$, the host worker of $G_i$ will be idle at the first superstep. 
In the subsequent supersteps, the designated messages $M_i$ to subgraph $G_i$ contain \emph{(v.id, dist(s,v))} pairs, and the distances are updated during the SBS. During the following \emph{Compute} phase, the updated frontier vertices are pushed into the \emph{min-priority queue} to propagate the external distance change to the entire subgraph.  
The $\min$ function is adopted as the \emph{Aggregate} operator for data coherence across subgraphs as part of SBS. 
\subsection{PageRank}\label{sec:pr}
PageRank(PR), an algorithm to rank every web page in a web graph.
Let vertex $u$ represent a web page and edge $(u, v)$ represent a hyperlink from $u$ to $v$.
With the notation defined in Section ~\ref{sec:back}, we have: 
\begin{equation}\label{eq:pr}
PR^{n}(u) = \alpha\sum\limits_{v \in N^{in}_u}{\frac{PR^{n-1}(v)}{|N^{out}_v|}} + \frac{1-\alpha}{N}
\end{equation}
supposing $PR^n(u)$ is the PageRank value for vertex $u$ in the $nth$ iteration. $\alpha$ is a damping factor which is usually set as $0.85$.
Note that Equation ~\ref{eq:pr} is a classic synchronous PageRank algorithm. Instead, in SVHM we implement the PageRank algorithm following an asynchronous accumulative approach proposed in ~\cite{zhang2012accelerate}.
We refer to $\delta_u$ as the accumulator of the received updates of $u$. Similar to the vertex-centric program, we loop over the \emph{"active"} vertices in the same partition and update their PageRank with $\delta_u$ at each superstep, $\alpha{\frac{\delta_u}{|N^{out}_u|}}$ is sent to $u's$ adjacent vertices, then reseting $\delta_u$ to $0$. The vertex $u$ will be identified as active if $\delta_u > 0$. In SVHM, each vertex could be activated by its neighbors in the same subgraph or by external updates from its remote replicas by \emph{SBS}. 
As the end of each superstep, $<u, \delta u>$ will be added to the SBS message vector if $u$ is a frontier vertex.
The SBS \textbf{Aggregate} operator is set to be $sum$ function.
\subsection{Graph Simulation}
Graph $G(V,E)$ matches a pattern graph $Q(V_Q,E_Q)$ via graph simulation ~\cite{henzinger1995computing}.
if there is a binary relation $R \subseteq V_Q \times V$ such that
    (a) For each vertex $u \in V_Q$, there exists a vertex $v \in V$ satisfies
    $(u,v) \in R$ and $u.label = v.label$,
    (b) For each pair $(u,v) \in R$, considering edge $(u, u')$
    in $E_Q$, a vertex $v' \in V$ can be found such that $(v, v') \in E$ and $(u', v') \in R$.
Our implementation is based on the following idea:
(1) construct an initial relation $R_0$ according to condition (a);
(2) prune $R_0$ iteratively until it satisfies condition (b).
Algorithm ~\ref{alg:graphsimulation} illustrates how the graph simulation algorithm is implemented in the \textbf{SVHM} abstraction.
At the first superstep, we assume $v \in sim(u)$ if $u.label = v.label$.
Let $R_0$ be the initial relation corresponding to $sim$ at the first superstep,
$R_0$ is a superset of the final $R_m$.
In the following supersteps, we prune $sim$ and $post$ under condition(b).
With $post(v)$, we can access $v$'s successor vertices's matching results easily.
If $post(v)[u'] = 0$ with $v \in sim(u)$ and $(u, u') \in E_Q$, $v$ will be removed from $sim(u)$.
$<v, \Delta post(v)>$ will be added to SBS message vector if $\Delta post(v)$ is non-zero during the superstep.
Given two message pairs of $<v, \Delta post_1(v)>$ and $<v, \Delta post_2(v)>$, the \emph{Aggregate} operator traverses
vertex $u$ in $\Delta post_1(v) \cup \Delta post_2(v)$, and calculates the result \\$\Delta post(v)[u] = \Delta post_1(v)[u] + \Delta post_2(v)[u]$.
\begin{algorithm}
	\caption{Graph Simulation Compute function}
	\label{alg:graphsimulation}
	\DontPrintSemicolon
	\KwIn{Subgraph $G$, Pattern graph $Q$, Updated pairs by Global Sync $M$}
	\KwOut{Modified subgraph $G'$, Changed data pairs $\Delta D$}
	
	Vector $vec \gets \emptyset$\;
	Simulation set $sim \gets \emptyset$, Post map $post \gets \emptyset$\;
	
	\If{$getSuperstep = 0$}{
		\For{$u \in V_Q$, $v \in V$}{
			\If{$u.label = v.label$}{
				$sim(u).add(v)$\;
				\For{$v' \in N^{in}_v$}{
					$post(v')[u] \gets post(v')[u] + 1$\;
				}
			}
		}
		\For{$v \in V$} {
			$vec.addPairToVector(v.id, post(v))$\;
		}
	}
	\Else {
		$tempPost \gets \emptyset$\;
		\For{$v \in M$}{
			\For{$u \in V_Q$}{
				\If{$v \in sim(u)$ and exist vertex $u' \in N^{out}_u$ such that $post(v)[u'] = 0$}{
					$sim(u).remove(v)$\;
					\For{$v' \in N^{in}_v$}{
						$tempPost(v')[u] \gets tempPost(v')[u] - 1$\;
						$post(v')[u] \gets post(v')[u] - 1$\;
					}
				}
			}
		}
		\For{$v \in tempPost$} {
			$vec.addPairToVector(v.id, tempPost(v))$\;
		}
	}
\end{algorithm}

\section{Experimental Evaluation}\label{sec:experiments}
In this section, we evaluate the performance of DRONE from different perspectives. We carry out performance comparisons with representative frameworks of \emph{vertex-centric} (Giraph, Graphlab) and \emph{subgraph-centric}  (Blogel). Four graph algorithms were implemented for the performance evaluation:
\emph{Single Source Shortest Path}, \emph{PageRank}, \emph{Connected Component}, \emph{Graph Simulation}. 
In addition to cross-system comparison, for a fair comparison of SVHM model and traditional edge-cut based subgraph-centric model (e.g. Giraph++), we also made DRONE support the edge-cut counterpart.
 To distinguish the two modes in the following sections, we identify vertex-cut model (i.e. SVHM) as DRONE-VC and edge-cut mode as DRONE-EC. Parmetis~\cite{karypis1997parmetis} is chosen as the default graph partitioner of DRONE-EC. However, in the case that Parmetis failed to partition when the graphs were too big for it to handle, we resorted to the random hashing (i.e. randomly assigning vertices to subgraphs) as the partitioner of DRONE-EC.  
Weak scaling capability of DRONE is tested on a HPC system with thousands of processes, and we present the result at the end of this section. 
Aside from framework evaluation, we also show the effectiveness of the partitioning strategy with the metrics that are defined in Section ~\ref{sec:distributedGraphPlacement}.

\subsection{Experimental Setup}
\begin{table*}
\centering
\begin{tabular}{@{}llllllll@{}}\toprule
Datasets &Type & $|V|$ & $|E|$ & Diameter & Average Degree & Max InDegree & Max OutDegree \\ \midrule
LiveJournal & Directed & 4,847,571 & 68,993,773 & 17 & 21.0 & 13,906 & 20,293 \\ 
USARoad & Undirected & 23,947,347 & 58,333,344 & 8,264 & 2.44 & 9 & 9\\ 
WebBase & Directed & 118,142,155 & 1,019,903,190 & 37 & 8.63 & 816,127 & 3,841\\ 
\bottomrule
\end{tabular}
\caption{Stats of Real-World DataSets}
\label{tab:datasets}
\end{table*}

Our in-house test bed is a 4-node system, each node consists of 8 Intel Xeon E7-8830 2.13GHz CPU total of 64 cores and 1TB memory. The system is connected via 1GigE Ethernet. For the cross-system comparison experiments, 
 we chose three open datasets that are the largest available to us. They are listed in Table 
 ~\ref{tab:datasets}: (1)social network: \emph{LiveJournal} ~\cite{LiveJournal}; (2)web graph: \emph{WebBase} ~\cite{WebBase} (3)road network: \emph{USARoad} ~\cite{USARoad}. Among these datasets, \emph{LiveJournal} and {WebBase} are power-law graphs, \emph{USARoad} has a large diameter(6000+ hops). To evaluate the weak scaling of DRONE, we chose larger synthetic \emph{power-law} graphs produced by the kronecker generator described in ~\cite{leskovec2010kronecker}, which is a power-law graph generator adopted by the \emph{Graph 500} benchmarks ~\cite{Graph500}. Weak scaling  experiments were carried out on \emph{"era"},  which has 270 nodes, each node with 24 processors (2 Intel Xeon E5-2680 V3 2.5GHz CPU), 256GB RAM, and a 100Gb EDR Infiniband interconnect.   
\begin{table*}[htbp]
\begin{tabular}{@{}llllll@{}}\toprule
& & \multicolumn{2}{c}{Imbalance} & \multicolumn{2}{c}{Replication Factor} \\
\cmidrule{3-4}  \cmidrule{5-6} 
DataSets & \#Subgraphs & Random Hash & Canonical Degree Based Hash& Random Hash & Canonical Degree Based Hash\\ \midrule
LiveJournal & 4   &  1.007 & 1.006 & 2.4691 & 2.41677\\ 
WebBase    & 32 & 1.03 & 1.02 & 6.29 & 6.0\\ 
\bottomrule
\end{tabular}
\caption{Partitioning metrics comparison of Random Hash and Canonical Degree-Based Hash over power-law graphs}
\label{tab:partitionercompare}
\vspace{-5pt}
\end{table*}

\begin{figure}[h]
\centering
\begin{minipage}{0.9\linewidth}
  \centerline{\includegraphics[width=1\textwidth]{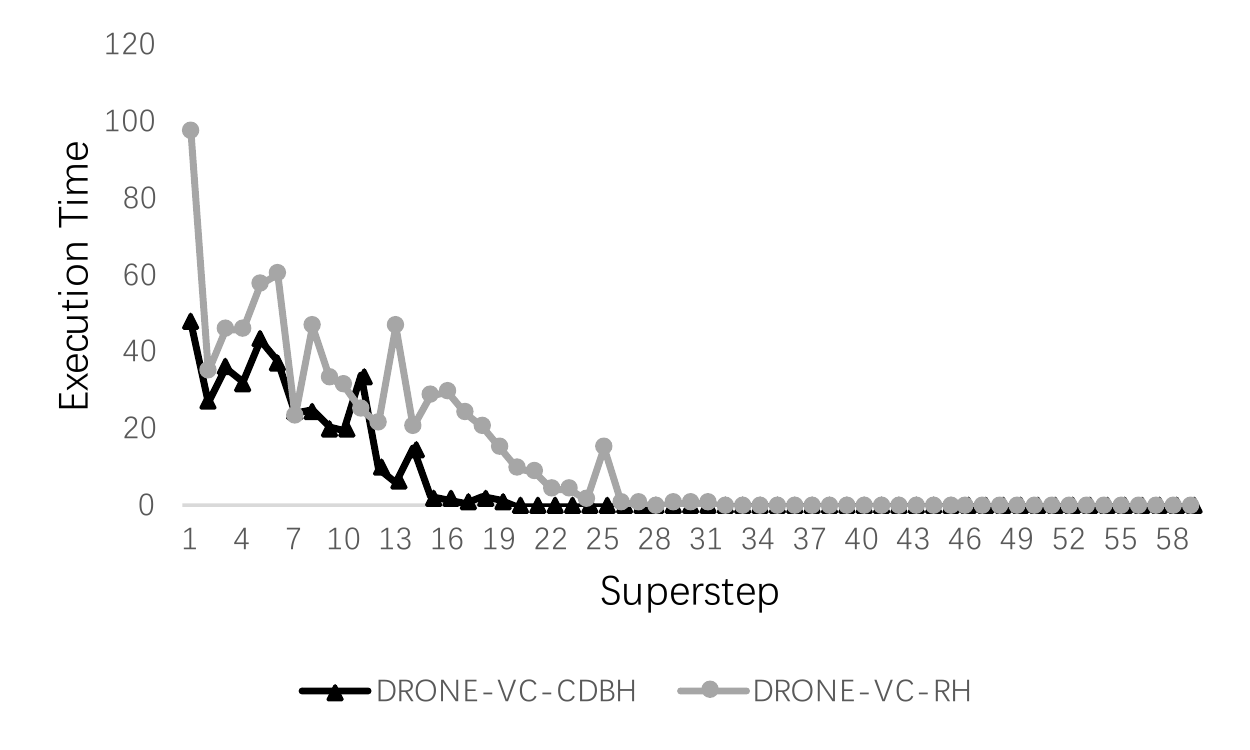}}
  \centerline{(a) Excution Time}
  \centerline{}
\end{minipage}
\vfill
\begin{minipage}{0.9\linewidth}
  \centerline{\includegraphics[width=1\textwidth]{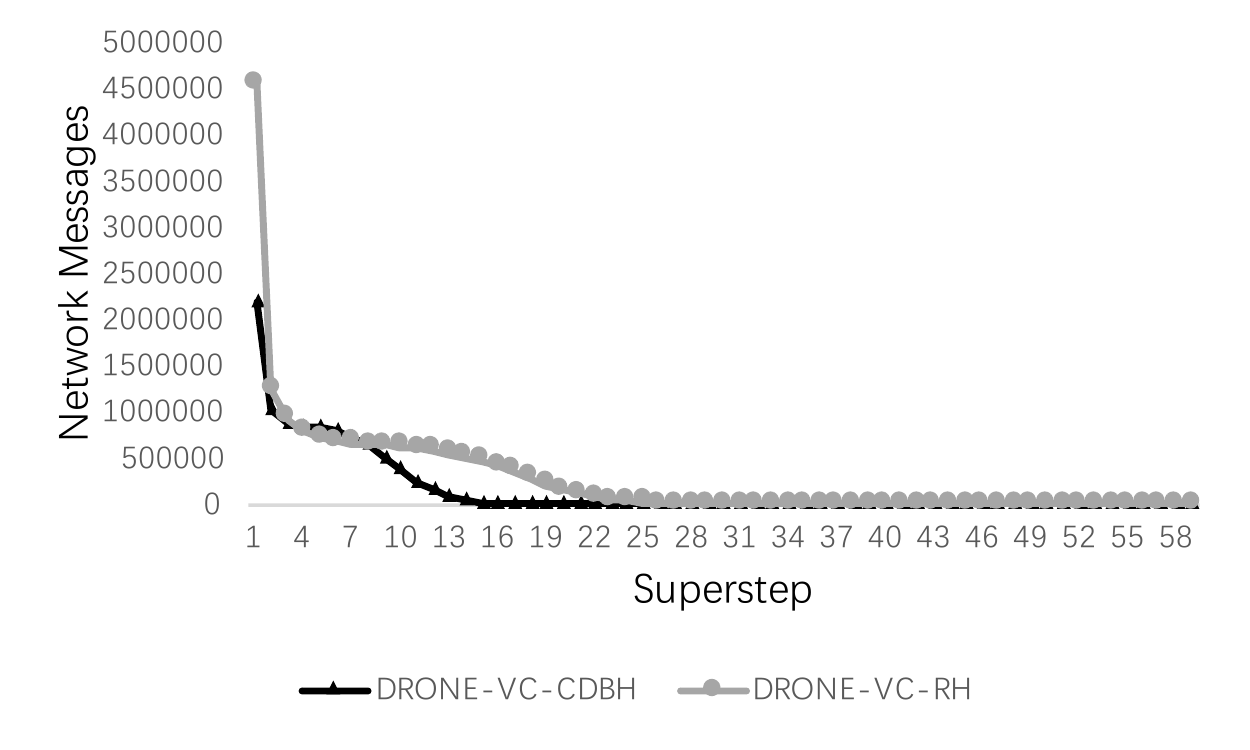}}
  \centerline{(b) Network Messages}
  \centerline{}
  \vspace{-10pt}
\end{minipage}
\caption{Partitioner performance comparison over CC execution time and network messages per superstep on WebBase graph}
\label{fig:partitionercompare}
\vspace{-15pt}
\end{figure}  

\subsection{Performance Comparison of Partitioner}\label{sec:partitionercompare}

Table ~\ref{tab:partitionercompare} shows the partitioning metrics comparison of imbalance and replication factors between Random Hashing (RH) and Canonical Degree Based Hashing (CDBH) vertex-cut partitioners. We can see that both vertex-cut partitioner divide graph evenly. The imbalance factors are approximate to 1 with difference less than $1\%$. 
Regarding the replication factor, \emph{CDBH} is slightly better than \emph{RH} over \emph{Webbase} and \emph{LiveJournal}, approximately $2.1\%$ and $4.8\%$ respectively.  
However, despite the marginal difference shown above, the CC algorithm achieves 1.96 times speedup(390 sec vs. 764 sec) with CDBH on \emph{webbase} dataset. There is 1.68 times improvement of network volume, i.e. less number of \emph{(key, value)} pairs exchanged across workers ($9,556,341$ vs. $16,121,171$). In addition, CC converges faster with CDBH, taking half the number of supersteps (508 vs. 1096) compared with RH. 
Figure ~\ref{fig:partitionercompare} shows the execution time and the number of key-value messages per superstep, and variations of the first 100 supersteps are displayed.  
 
\subsection{Systematic Performance Comparison}\label{sec:systemcompare}
In this section, we present a strong scaling performance comparison of DRONE, Giraph, Graphlab (Powergraph), and Blogel. We run CC and Sim over LiveJournal, PR over WebBase, SSSP over USARoad and Webbase. Due to the size of datasets, we limit our experiments to 24 workers. Although we carried out experiments for Spark GraphX, its performance was not comparable to the peers, so we decided not to include it in the following discussion.
\begin{figure*}
\centering
\begin{minipage}{0.24\linewidth}
  \centerline{\includegraphics[width=1\textwidth]{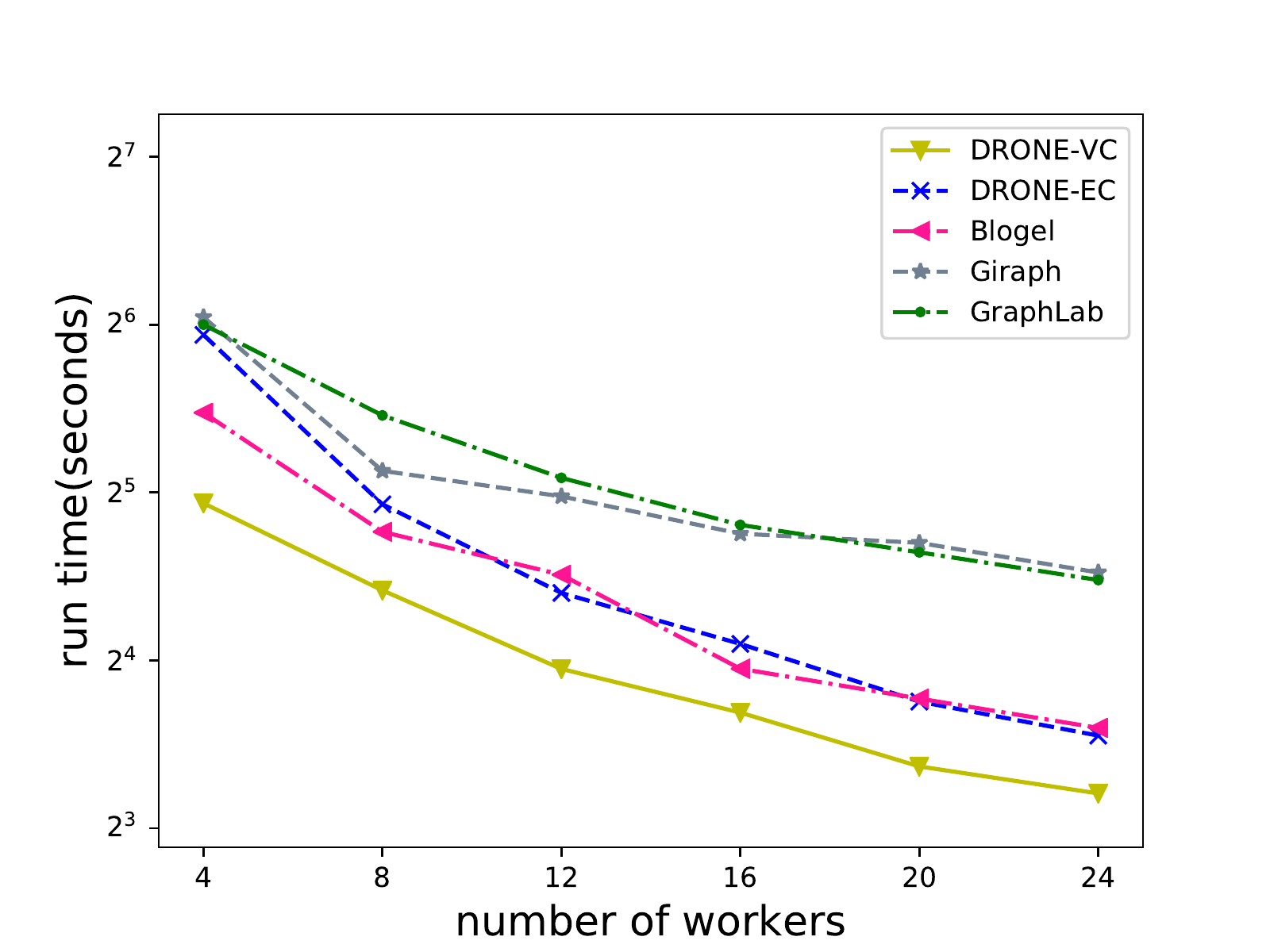}}
  \centerline{(a) CC}
\end{minipage}
\hfill
\begin{minipage}{0.24\linewidth}
  \centerline{\includegraphics[width=1\textwidth]{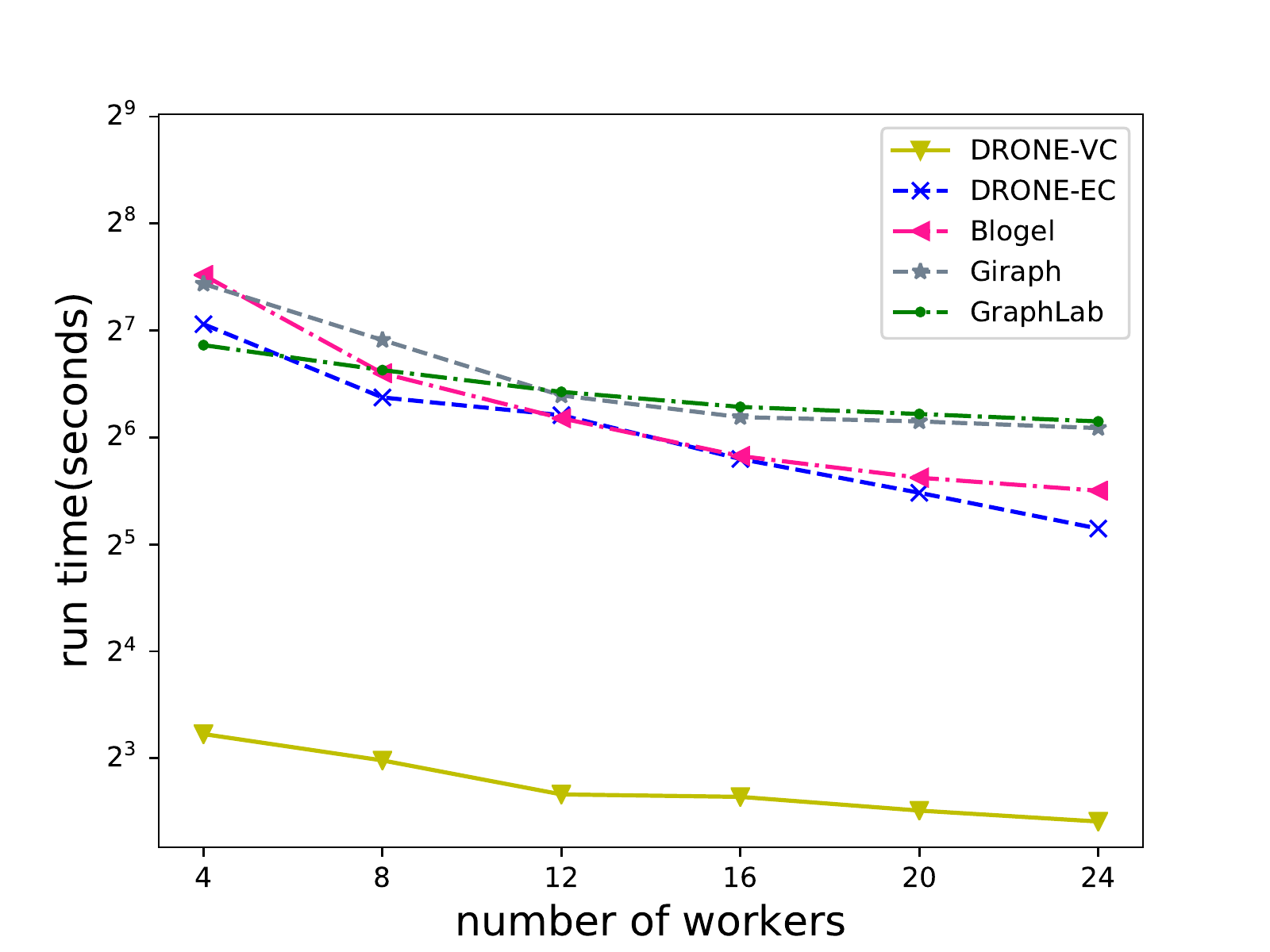}}
  \centerline{(b) GSim}
\end{minipage}
\hfill
\begin{minipage}{0.24\linewidth}
  \centerline{\includegraphics[width=1\textwidth]{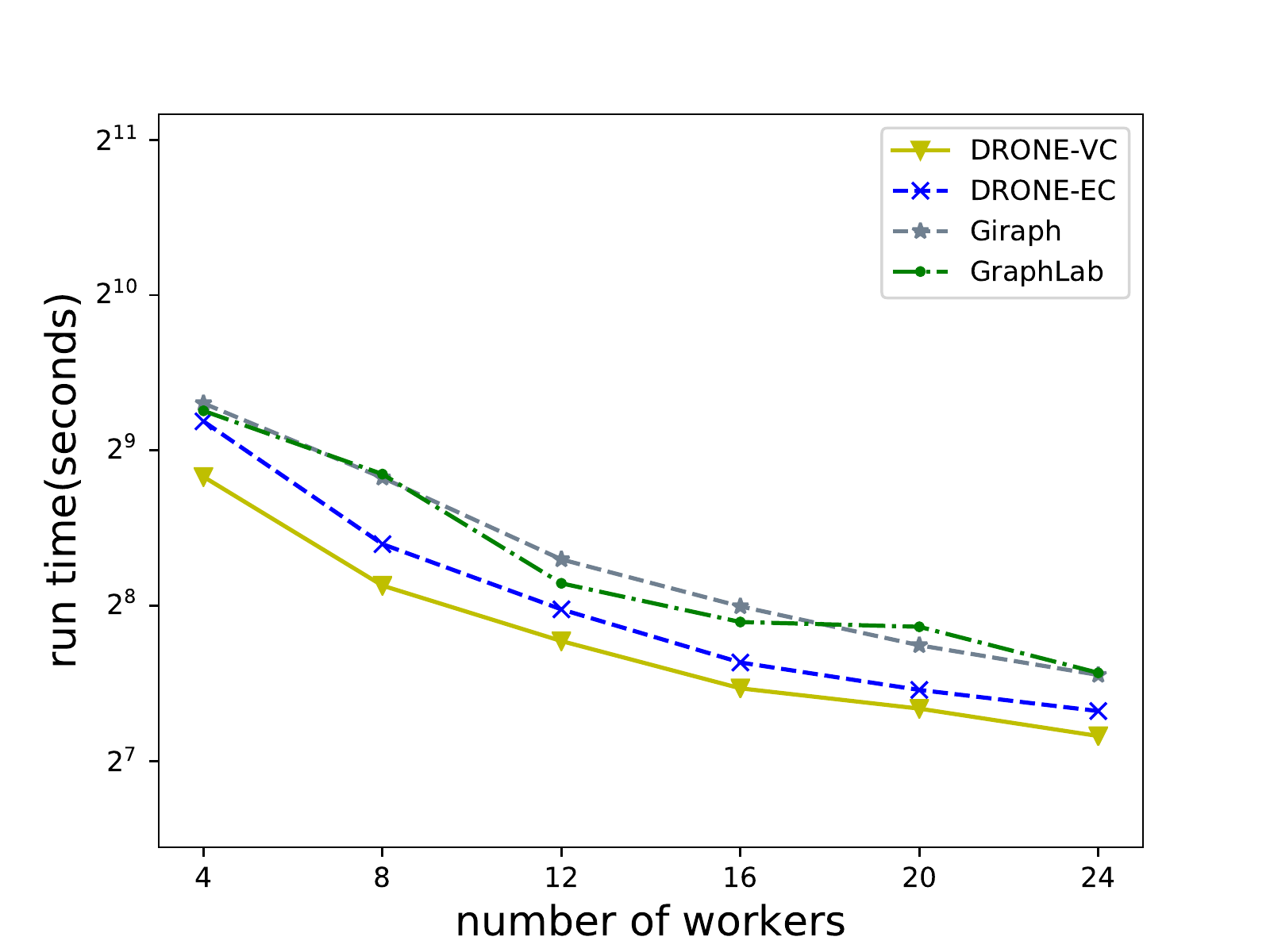}}
  \centerline{(c) PR}
\end{minipage}
\hfill
\begin{minipage}{0.24\linewidth}
  \centerline{\includegraphics[width=1\textwidth]{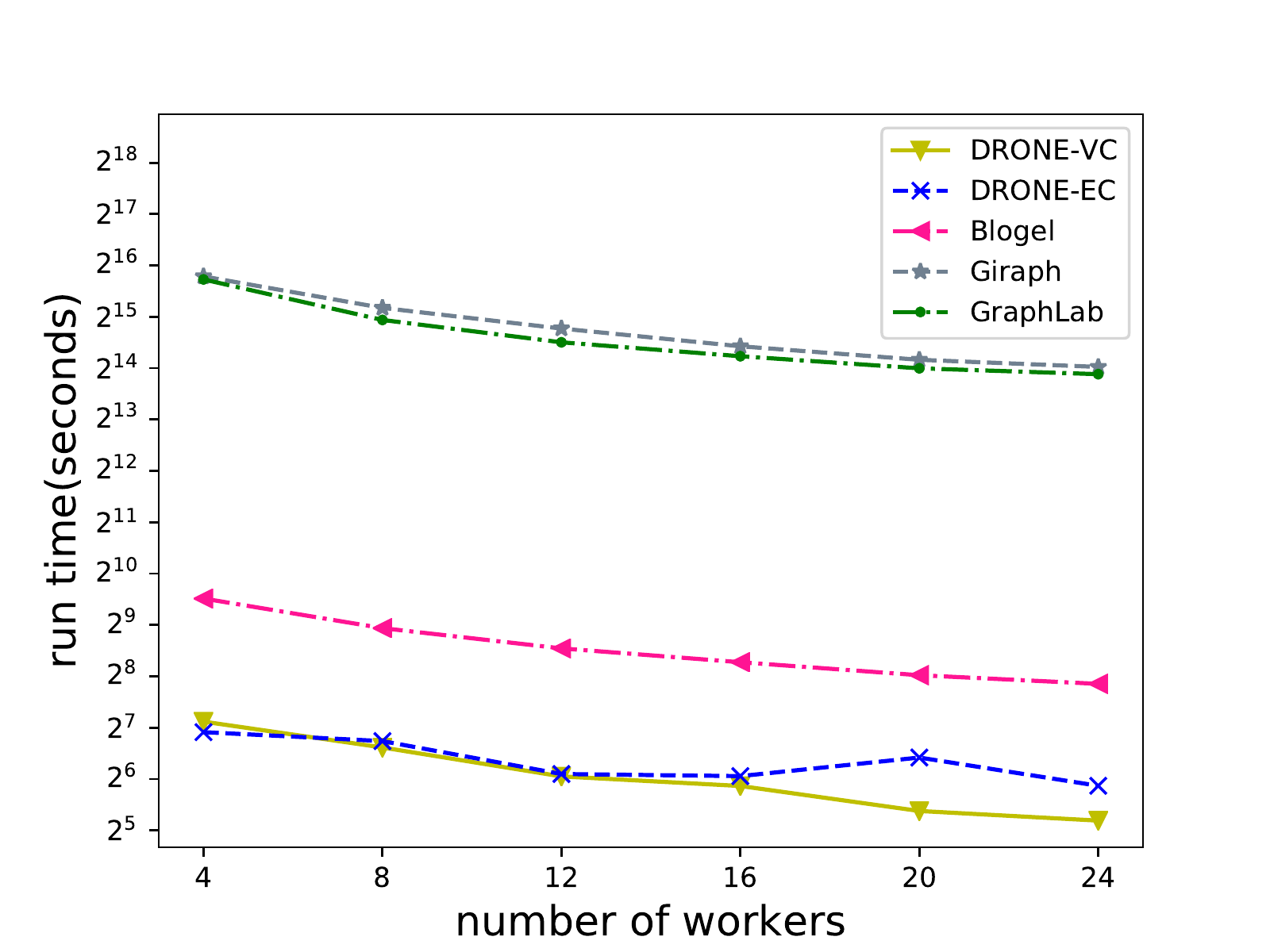}}
  \centerline{(d) SSSP}
\end{minipage}
\caption{Cross system strong scaling performance comparison}
\label{fig:systemcompare}
\end{figure*}

Figure ~\ref{fig:systemcompare} shows the strong scaling of four algorithms, namely CC, SSSP, GSim, and PR. In general, all systems exhibit a certain degree of scalability up to 24 workers. GraphLab and Giraph perform similarly on all algorithms. DRONE, which is identified as DRONE-VC, outperforms its peers in the aspects of execution time and scalability. In some cases, it is able to achieve a couple of order of magnitudes performance improvements. 
\paragraph*{CC}
Figure ~\ref{fig:systemcompare}(a) shows CC performance over \emph{LiveJournal}. DRONE-VC (i.e. SVHM implementation) outperforms DRONE-EC, Blogel, Giraph and Graphlab. DRONE-VC is by average 2.4 times faster than Giraph and Graphlab, both systems are vertex-centric frameworks. DRONE-VC is able to minimize the number of messages and achieve improvement of 170X compared to Giraph and Graphlab. We remark that for Blogel, it employs an multi-source BFS based partitioner to detect the connected components in the partitioning phase, ensures that each block is \emph{connected}. In the subsequent CC computing phase, it only merges small connected components into a bigger one without performing BFS. For a fare comparison, we added the precomputing time in partitioning phase to Blogel's CC running time. DRONE-VC outperforms Blogel by average 1.3 times,
this is because CC is a simple graph traversal algorithm, the running time is dominated by the communication overhead. 
For instance, when n = 24, DRONE-VC only transmits $0.5\%$ of the subgraph-level messages transmitted by Blogel. DRONE-VC transmits 3.5 times fewer \emph{(key, value)} pairs resulting in 1.4 times faster execution time than DRONE-EC.
\paragraph*{GSim}
Figure ~\ref{fig:systemcompare}(b) shows the performance comparison of the GSim algorithm. DRONE-VC is on average 6.7x, 8.5x, 12.8x, 13.4x faster than DRONE-EC, Blogel, Giraph,and Graphlab. 
DRONE exhibits superior performance over its peers, which demonstrates that the SVHM is suitable for algorithms like graph pattern matching.
\paragraph*{PR}
Figure ~\ref{fig:systemcompare} (c) shows the performance comparison of the PR algorithm. DRONE-VC performs better than other systems. Blogel is excluded from the comparison because its PR implementation is an optimized PageRank algorithm, The interested reader is refered to ~\cite{kamvar2003exploiting} for detailed information. 
We can see that DRONE-VC and DRONE-EC are significantly faster than vertex-centric systems (i.e. Giraph and Graphlab) due to less communication. We would like to point out that both subgraph-centric and vertex-centric models take a similar number of supersteps. 
\paragraph*{SSSP}
\begin{figure}
  \centerline{\includegraphics[width=0.9\linewidth]{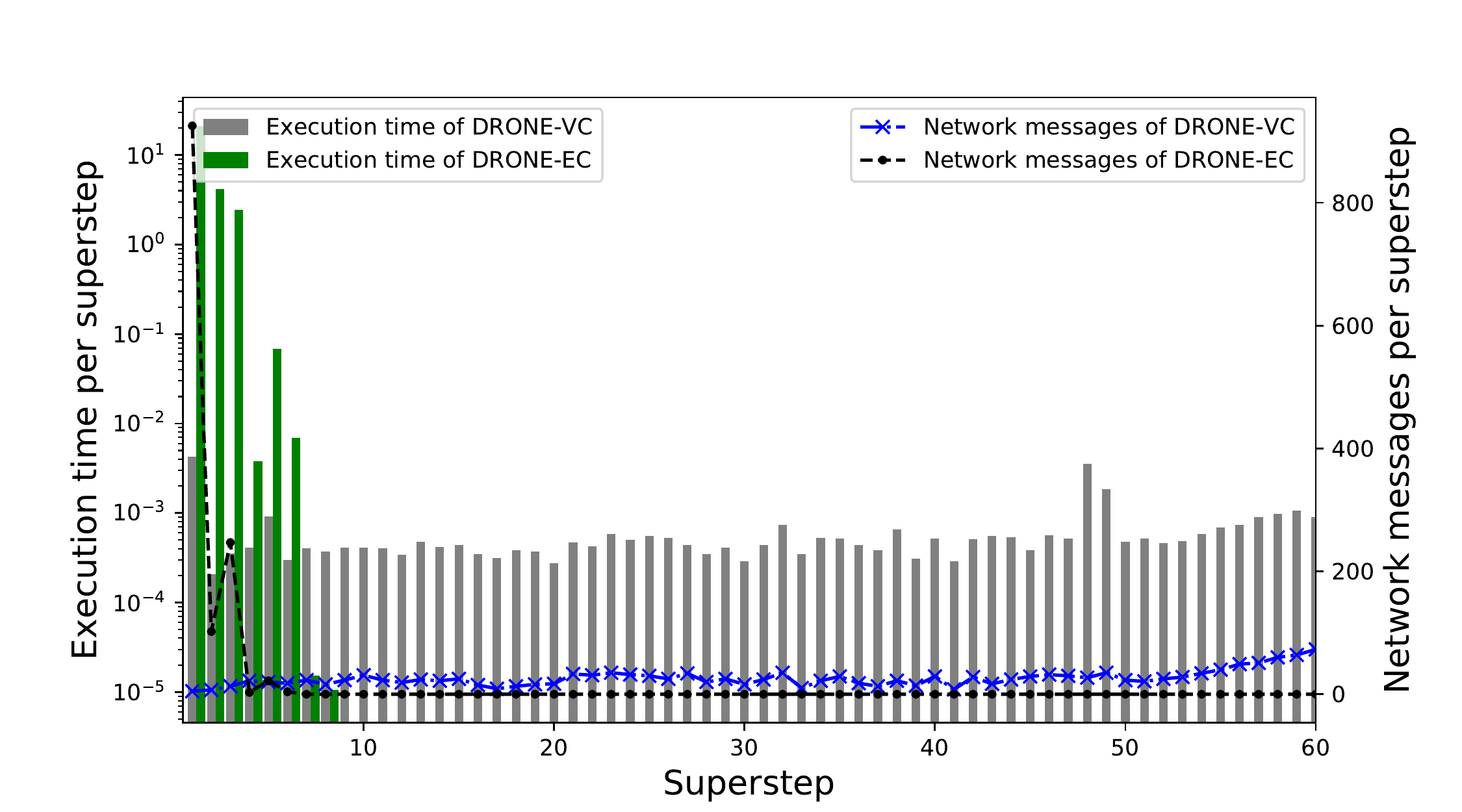}}
\caption{ SSSP runtime and messages per superstep for DRONE-EC and DRONE-VC}
\label{fig:ssspcomp}
\end{figure}
Figure ~\ref{fig:systemcompare}(d)  shows the performance of SSSP over \emph{USARoad}. DRONE-VC significantly outperforms the other systems. With 24 workers, DRONE-VC is 455x, 413x, 6.3x and 1.6x faster than Giraph, Graphlab, Blogel and DRONE-EC, respectively. Subgraph-centric systems(i.e DRONE, Blogel) take significantly fewer supersteps than vertex-centric systems(i.e Giraph, Graphlab). 
 USARoad is not a power-law graph, which has relatively even degree distributions. 
Metis, default partitioner of DRONE-EC, tries to minimize the edge cuts and splits graphs with balanced number of vertices. On the contrary, CDBH evenly distributes edges to subgraphs, and acts like RH on USARoad. It is interesting to see that DRONE-EC completes with fewer supersteps, but take longer time to finish. Figure ~\ref{fig:ssspcomp} shows the runtime and messages per superstep for DRONE-VC and DRONE-EC. We can see that DRONE-EC spends much more time during first few steps and sending more messages than DRONE-VC. Although DRONE-VC takes more steps to complete, it outperforms DRONE-EC with a 1.6x speedup. It is evident that the \emph{vertex-cut} partitioning strategy of the SVHM abstraction ensures a balanced workload and good performance.

\begin{figure}[h]
\centering
\begin{minipage}{0.8\linewidth}
  \centerline{\includegraphics[width=1\textwidth]{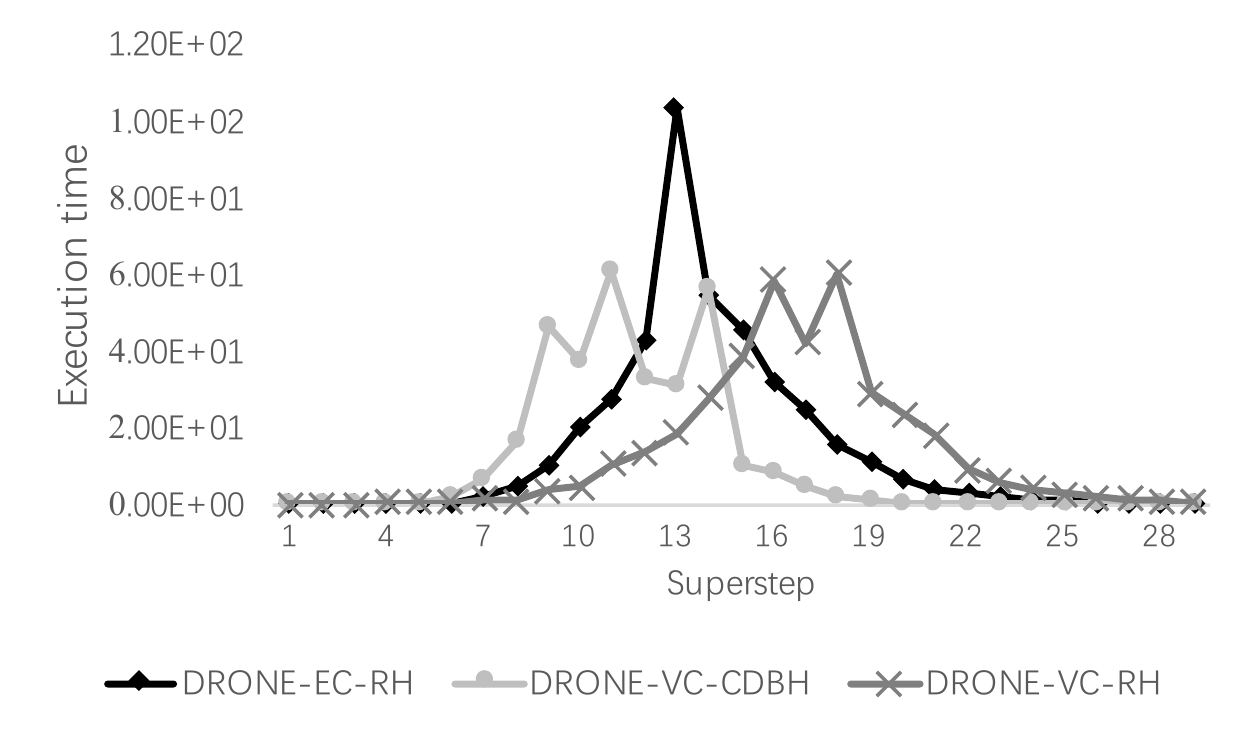}}
  \centerline{(a) Execution Time}
  \centerline{}
\end{minipage}
\vfill
\begin{minipage}{0.8\linewidth}
  \centerline{\includegraphics[width=1\textwidth]{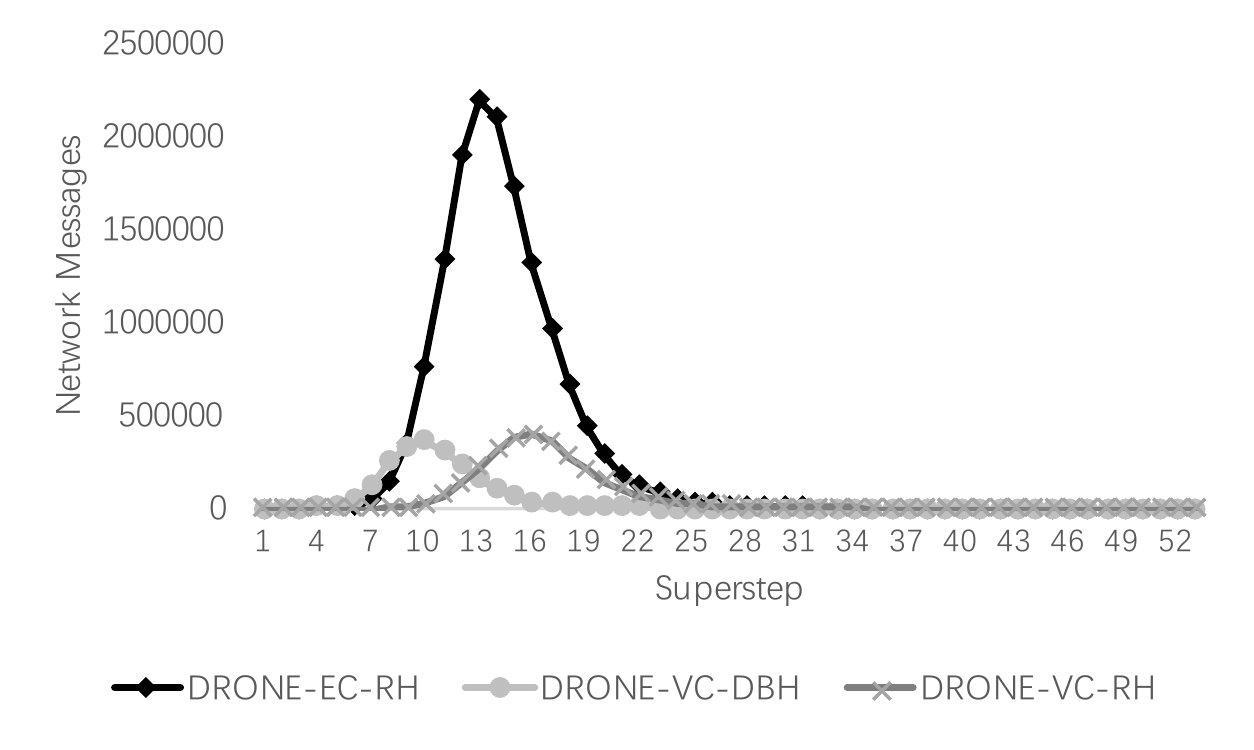}}
  \centerline{(b) Network Messages}
  \centerline{}
\end{minipage}
\caption{ SSSP performance per superstep for DRONE-EC-RH, DRONE-VC-RH, DRONE-VC-CDBH over Webbase dataset}
\label{fig:runtime_network_message_sssp_webbase}
\vspace{-15pt}
\end{figure}

We further compare SSSP performance of DRONE-EC and DRONE-VC on a larger dataset, Webbase, which is a power-law graph. For a fair comparison, we tested three variations: DRONE-EC (random hashing edge-cut), DRONE-VC (random hashing vertex-cut), and DRONE-VC (CDBH). Figure ~\ref{fig:runtime_network_message_sssp_webbase} shows execution time and network messages per superstep.
\begin{table}  
\centering  
\begin{tabular}{@{}lll@{}}\toprule
 & Execution time(s) & Supersteps  \\ \midrule
DRONE-VC-CDBH & 316 & 261\\ 
DRONE-VC-RH & 378 & 535\\ 
DRONE-EC-RH & 413 & 607\\ 
\bottomrule
\vspace{-15pt}
\end{tabular}   
\caption{SSSP performance comparison of DRONE variants on Webbase} 
\label{tab:sssp-drone}  
\end{table}  
Although both figures show similar trends for the three variations, DRONE-VC-CDBH achieves the best performance, see a detail in Table ~\ref{tab:sssp-drone}.  We would like to point out that the number of network messages transmitted by DRONE-VC-RH is only $19\%$ of DRONE-EC-RH($2,097,868$ vs. $14,761,958$), and DRONE-VC-RH only takes $42\%$ the numbers of supersteps of DRONE-EC-RH.
\subsection{Breakdown of Execution Time}
\begin{figure}[ht]
\centering
\begin{minipage}{0.9\linewidth}
  \centerline{\includegraphics[width=1\textwidth]{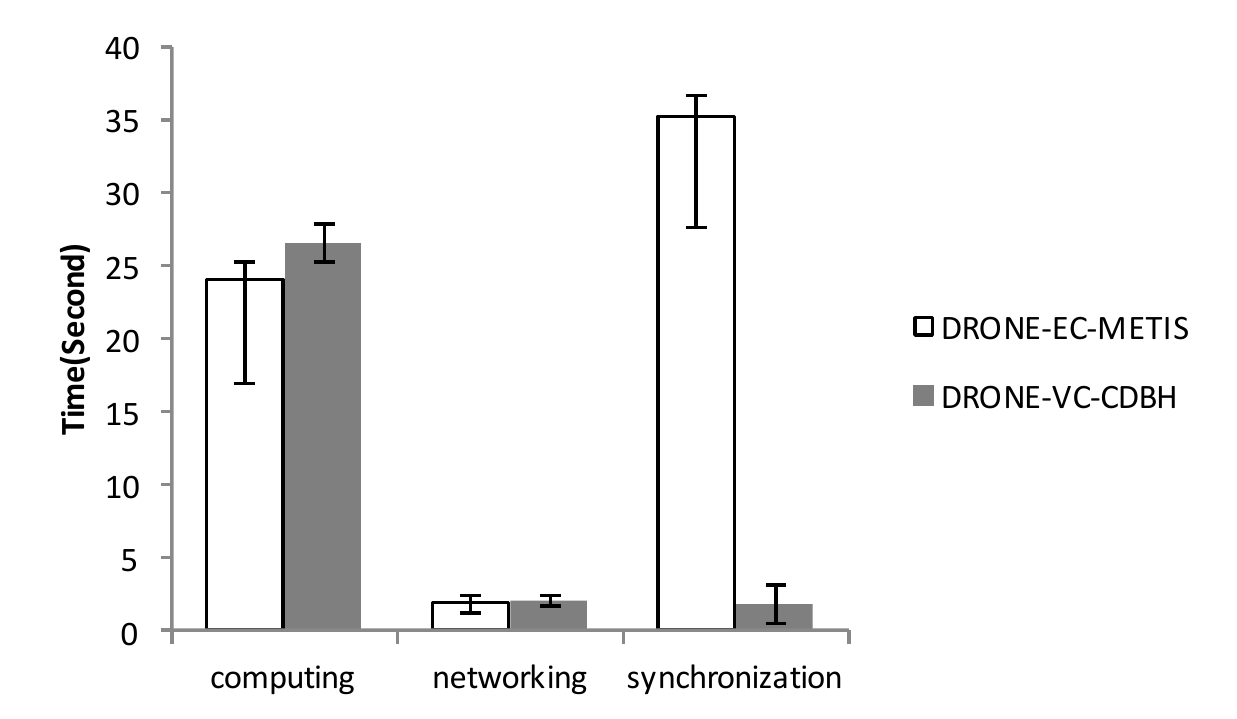}}
  \centerline{(a) Execution time break-down of CC over LiveJournal}
  \centerline{}
\end{minipage}
\vfill
\begin{minipage}{0.9\linewidth}
  \centerline{\includegraphics[width=1\textwidth]{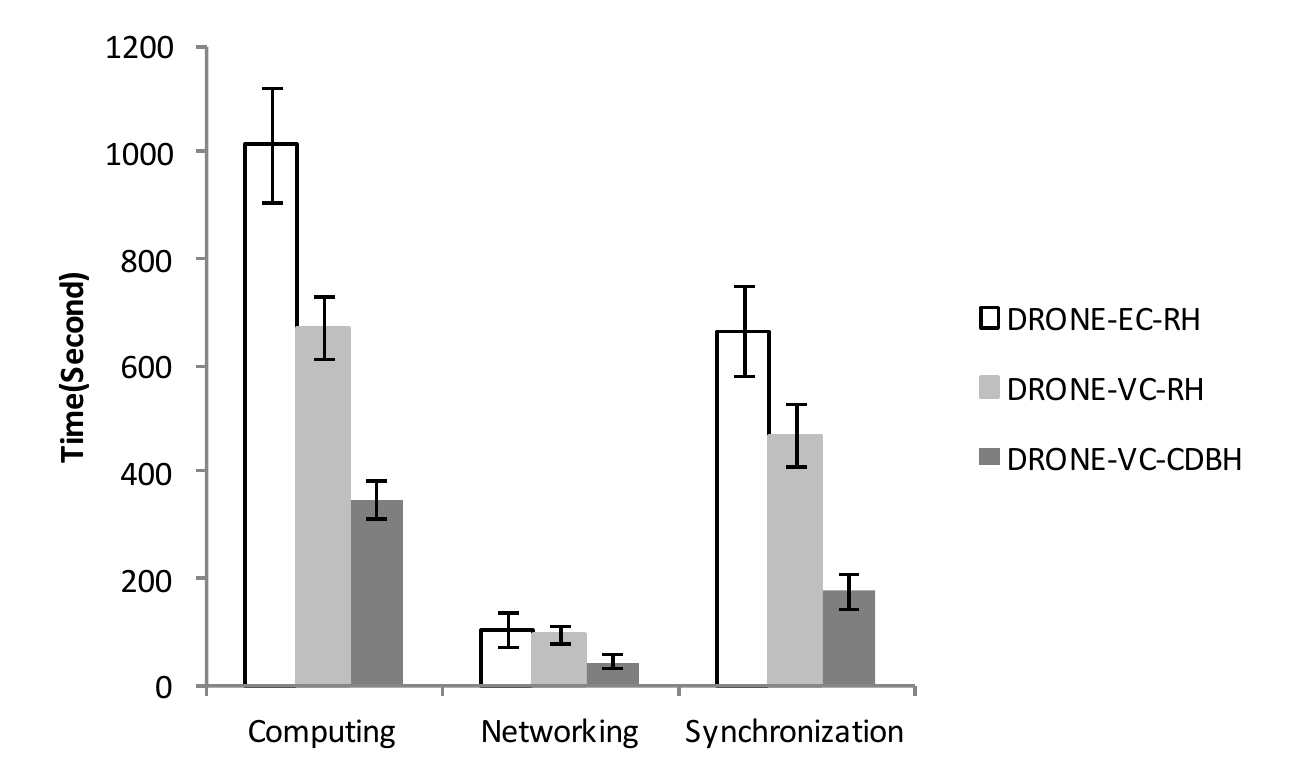}}
  \centerline{(b) Execution time break-down of CC over Webbase}
  \centerline{}
\end{minipage}
\vfill
\begin{minipage}{0.9\linewidth}
  \centerline{\includegraphics[width=1\textwidth]{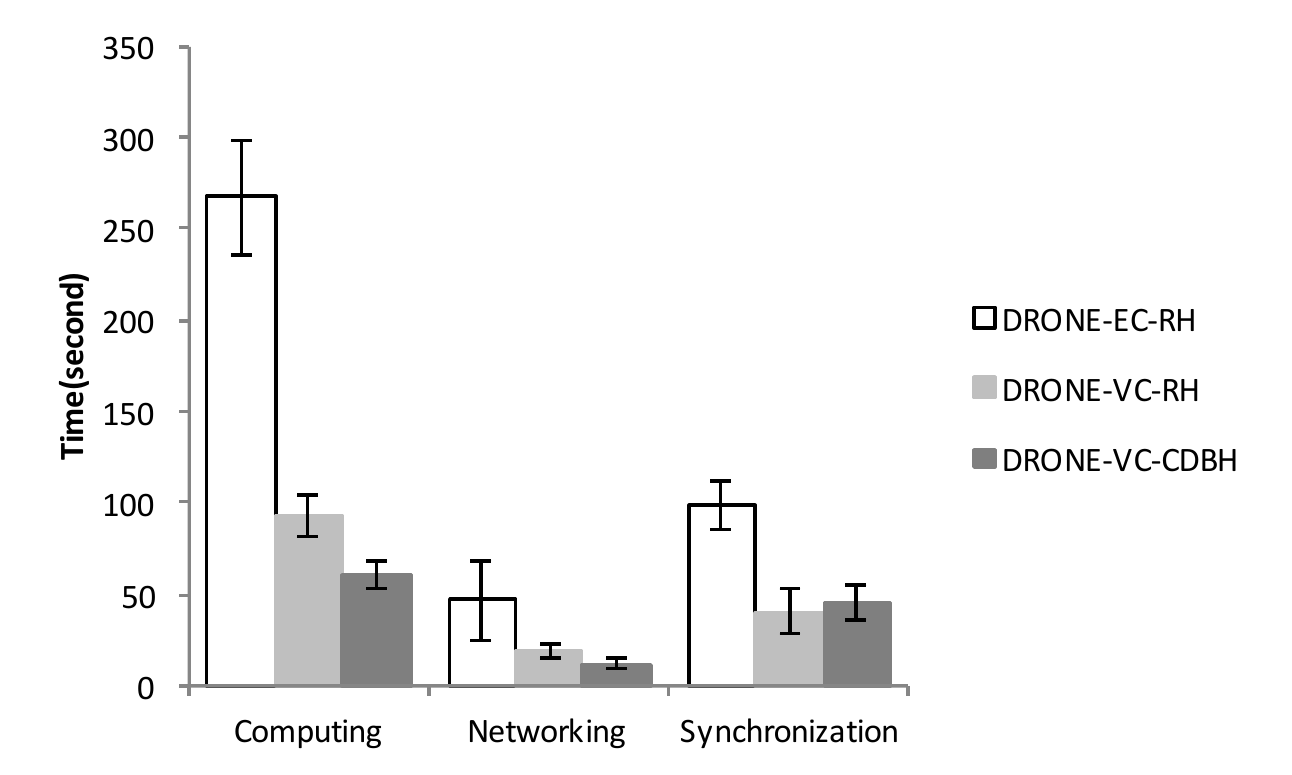}}
  \centerline{(c) Execution time break-down of SSSP over Webbase}
  \centerline{}
\end{minipage}
\caption{ Execution time breakdown}
\label{fig:breakdown}
\vspace{-15pt}
\end{figure}

In this section, we have a close look at the performance of the SVHM abstraction and the traditional edge-cut based subgraph-centric model. We breakdown the execution time into three parts: (1) Computing: the execution of user-defined graph algorithm; (2) Networking: The time of sending and receiving between workers. (3) Synchronization. 
In Figure ~\ref{fig:breakdown}, we plot min and max on each bar, which represents average value for a category. Note that higher deviation implies greater imbalance and lower average time implies better performance. Figure ~\ref{fig:breakdown}(a) reports the execution time breakdown of the CC algorithm under DRONE-EC and DRONE-VC mode over \emph{LiveJournal}. PARMETIS~\cite{karypis1997parmetis} is chosen to be the edge-cut partitioner for DRONE-EC. We can see that in terms of average computing and networking time, DRONE-EC and DRONE-VC have close performance. However, DRONE-EC suffers from performance degradation due to significant workload imbalance, as workers spend more time on synchronization. 
PARMTIS is too expensive and not competent to generate balanced workload partitions when dealing with graphs of highly skewed degree distribution.
Figure ~\ref{fig:breakdown}(b) and Figure ~\ref{fig:breakdown}(c) report the execution time breakdown of CC and SSSP over \emph{Webbase}, respectively. Because PARMETIS failed to partition \emph{Webbase}, we chose the Random Hashing(RH) partitioner for comparison.
We can see that DRONE-VC-CDBH performs significantly better for both algorithms in all categories. Although both utilized the same RH partitioning strategy, DRONE-VC-RH (vertex-cut partitioning) significantly outperforms DRONE-EC-RH (edge-cut partitioning). Specifically, for the SSSP algorithm, DRONE-VC-RH is 4.45x faster than DRONE-EC-RH in average computing time and only takes $40\%$ synchronization time of DRONE-EC-RH, which implies less workload imbalance. 

\subsection{Weak Scaling Experiment}
\begin{figure}[htbp]
\centering
\includegraphics[width=0.9\linewidth]{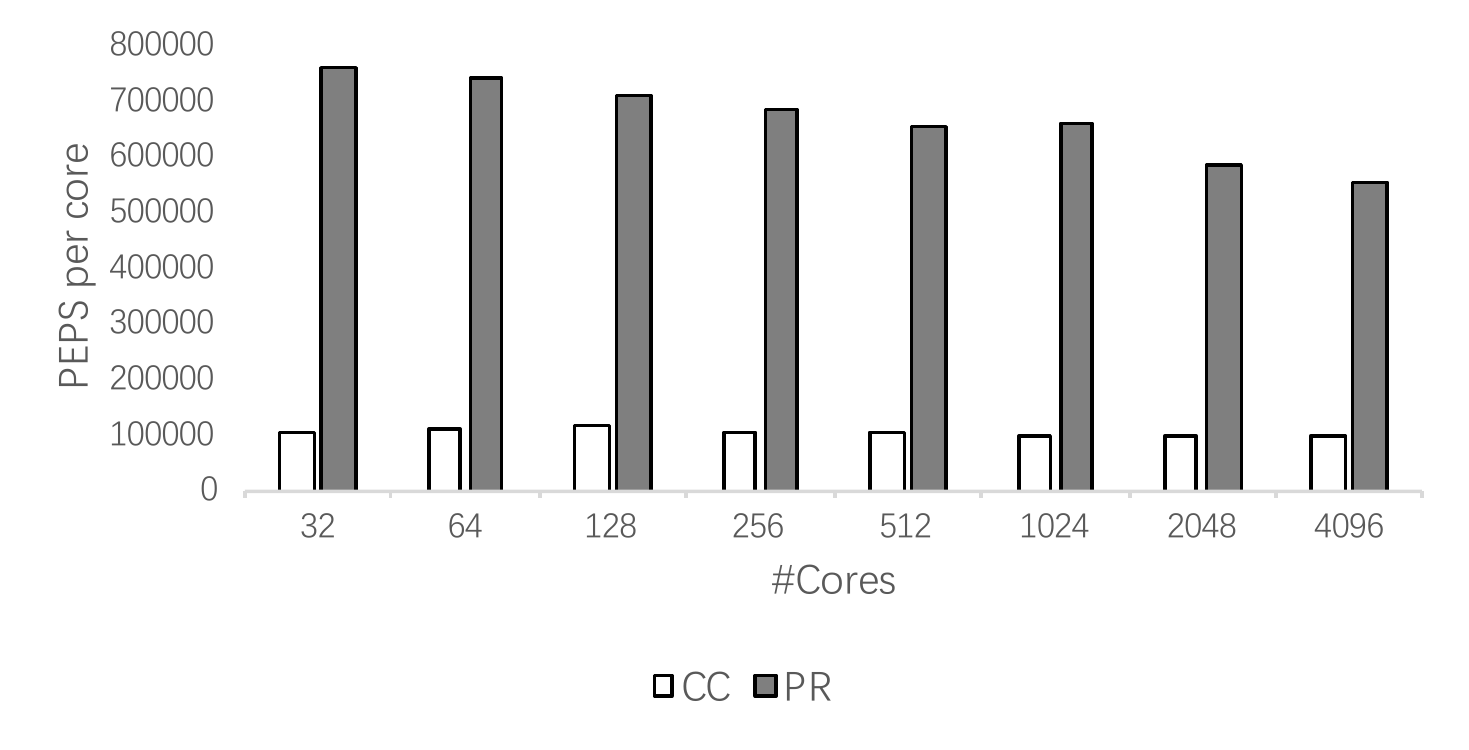}
\caption{Weak scaling of PR and CC on Kronecker graphs}
\label{fig:weakscaling}
\vspace{-15pt}
\end{figure}

In this section, we evaluated the weak scaling of DRONE. We executed two algorithms: CC and PR on a range of synthetic graphs created by the \emph{Kronecker} graph generator ~\cite{leskovec2010kronecker}. We use the same parameters from Graph500 for generating synthetic graphs. 
For our weak scaling experiments, we fix the average number of edges of each partition and generate Kronecker graphs with scales ranging from $2^{30}$ on 32 cores to $2^{37}$ on 4096 cores. The performance is measured by \emph{PEPS  (actual processed edges per second)}
~\cite{gb-tsinghua}. 
The average \emph{PEPS} per core (worker) is presented in Figure ~\ref{fig:weakscaling}. We can see that DRONE achieves suboptimal weak scalability to process large scale power-law graphs on HPC system. 

\section{Conclusion and Future Work}\label{sec:conclusion}
In the scenario of processing massive large scale graph-structure data, various graph-parallel models have emerged.  The \emph{subgraph-centric} model provides users with the opportunity to freely program against the subgraph, allowing users to reuse the existing sequential graph algorithms. Compared with the vertex-centric model, subgraph-centric may significantly improve efficiency for performing the algorithms that are topology-aware upon the well-partitioned graphs. However, the existing subgraph-centric frameworks default to edge-cut schemes when partitioning power-law graphs, is vulnerable to the workload imbalance, large inter-machine communication, and requires heavy preprocessing. We observed these defects in practice, and proposed the SVHM abstraction and implemented it in a system called DRONE. By exploiting the merits of vertex-cut to reduce communication and storage overhead, and achieve load balance, SVHM can efficiently process large scale power-law graphs. The API of DRONE is flexible and easy to program with, and users can express the algorithms with it. Our experiments show that DRONE is competitive with the state-of-art distributed graph processing frameworks

In the future, we would like to extend DRONE to asynchronous distributed graph processing, and explore graph algorithms for machine learning. 
\bibliographystyle{abbrv}


\begin{thebibliography}{10}

\bibitem{Giraph}
Apache: Giraph.
\newblock \url{http://giraph.apache.org}.

\bibitem{Graph500}
Graph500.
\newblock \url{https://graph500.org}.

\bibitem{LiveJournal}
Livejournal.
\newblock \url{http://snap.stanford.edu/data/com-LiveJournal.html}.

\bibitem{USARoad}
Usaroad.
\newblock \url{https://www.dis.uniroma1.it/challenge9/download.shtml}.

\bibitem{WebBase}
Webbase.
\newblock \url{https://law.di.unimi.it/webdata/webbase-2001/}.

\bibitem{bourse2014balanced}
F.~Bourse, M.~Lelarge, and M.~Vojnovic.
\newblock Balanced graph edge partition.
\newblock In {\em Proceedings of the 20th ACM SIGKDD international conference
  on Knowledge discovery and data mining}, pages 1456--1465. ACM, 2014.

\bibitem{ching2015one}
A.~Ching, S.~Edunov, M.~Kabiljo, D.~Logothetis, and S.~Muthukrishnan.
\newblock One trillion edges: Graph processing at facebook-scale.
\newblock {\em Proceedings of the VLDB Endowment}, 8(12):1804--1815, 2015.

\bibitem{faloutsos1999powergraph16}
M.~Faloutsos, P.~Faloutsos, and C.~Faloutsos.
\newblock On power-law relationships of the internet topology.
\newblock In {\em ACM SIGCOMM computer communication review}, volume~29, pages
  251--262. ACM, 1999.

\bibitem{fan2017parallelizing}
W.~Fan, J.~Xu, Y.~Wu, W.~Yu, J.~Jiang, Z.~Zheng, B.~Zhang, Y.~Cao, and C.~Tian.
\newblock Parallelizing sequential graph computations.
\newblock In {\em Proceedings of the 2017 ACM International Conference on
  Management of Data}, pages 495--510. ACM, 2017.

\bibitem{fredman1987fibonacci}
M.~L. Fredman and R.~E. Tarjan.
\newblock Fibonacci heaps and their uses in improved network optimization
  algorithms.
\newblock {\em Journal of the ACM (JACM)}, 34(3):596--615, 1987.

\bibitem{gonzalez2012powergraph}
J.~E. Gonzalez, Y.~Low, H.~Gu, D.~Bickson, and C.~Guestrin.
\newblock Powergraph: distributed graph-parallel computation on natural graphs.
\newblock In {\em OSDI}, volume~12, page~2, 2012.

\bibitem{guerrieri2014distributed}
A.~Guerrieri and A.~Montresor.
\newblock Distributed edge partitioning for graph processing.
\newblock {\em arXiv preprint arXiv:1403.6270}, 2014.

\bibitem{henzinger1995computing}
M.~R. Henzinger, T.~A. Henzinger, and P.~W. Kopke.
\newblock Computing simulations on finite and infinite graphs.
\newblock In {\em Foundations of Computer Science, 1995. Proceedings., 36th
  Annual Symposium on}, pages 453--462. IEEE, 1995.

\bibitem{DBLP:journals/tkde/KalavriVH18}
V.~Kalavri, V.~Vlassov, and S.~Haridi.
\newblock High-level programming abstractions for distributed graph processing.
\newblock {\em {IEEE} Trans. Knowl. Data Eng.}, 30(2):305--324, 2018.

\bibitem{kamvar2003exploiting}
S.~Kamvar, T.~Haveliwala, C.~Manning, and G.~Golub.
\newblock Exploiting the block structure of the web for computing pagerank.
\newblock Technical report, Stanford, 2003.

\bibitem{karypis1997parmetis}
G.~Karypis and V.~Kumar.
\newblock A coarse-grain parallel formulation of multilevel k-way graph
  partitioning algorithm.
\newblock In {\em PPSC}, 1997.

\bibitem{leskovec2010kronecker}
J.~Leskovec, D.~Chakrabarti, J.~Kleinberg, C.~Faloutsos, and Z.~Ghahramani.
\newblock Kronecker graphs: An approach to modeling networks.
\newblock {\em Journal of Machine Learning Research}, 11(Feb):985--1042, 2010.

\bibitem{li2014scaling}
M.~Li, D.~G. Andersen, J.~W. Park, A.~J. Smola, A.~Ahmed, V.~Josifovski,
  J.~Long, E.~J. Shekita, and B.-Y. Su.
\newblock Scaling distributed machine learning with the parameter server.
\newblock In {\em OSDI}, volume~14, pages 583--598, 2014.

\bibitem{gb-tsinghua}
H.~Lin, X.~Zhu, B.~Yu, X.~Tang, W.~Xue, W.~Chen, L.~Zhang, T.~Hoefler, X.~Ma,
  X.~Liu, W.~Zheng, and J.~Xu.
\newblock {ShenTu: Processing Multi-Trillion Edge Graphs on Millions of Cores
  in Seconds}.
\newblock In {\em Proceedings of the International Conference for High
  Performance Computing, Networking, Storage and Analysis (SC18) - Gordon Bell
  Award Finalist}. ACM, Nov. 2018.

\bibitem{low2012distributed}
Y.~Low, D.~Bickson, J.~Gonzalez, C.~Guestrin, A.~Kyrola, and J.~M. Hellerstein.
\newblock Distributed graphlab: a framework for machine learning and data
  mining in the cloud.
\newblock {\em Proceedings of the VLDB Endowment}, 5(8):716--727, 2012.

\bibitem{malewicz2010pregel}
G.~Malewicz, M.~H. Austern, A.~J. Bik, J.~C. Dehnert, I.~Horn, N.~Leiser, and
  G.~Czajkowski.
\newblock Pregel: a system for large-scale graph processing.
\newblock In {\em Proceedings of the 2010 ACM SIGMOD International Conference
  on Management of data}, pages 135--146. ACM, 2010.

\bibitem{mykhailenko2017distributed}
H.~Mykhailenko.
\newblock {\em Distributed edge partitioning}.
\newblock PhD thesis, Universit{\'e} C{\^o}te d'Azur, CNRS, I3S, France, 2017.

\bibitem{simmhan2014goffish}
Y.~Simmhan, A.~Kumbhare, C.~Wickramaarachchi, S.~Nagarkar, S.~Ravi,
  C.~Raghavendra, and V.~Prasanna.
\newblock Goffish: A sub-graph centric framework for large-scale graph
  analytics.
\newblock In {\em European Conference on Parallel Processing}, pages 451--462.
  Springer, 2014.

\bibitem{tian2013think}
Y.~Tian, A.~Balmin, S.~A. Corsten, S.~Tatikonda, and J.~McPherson.
\newblock From think like a vertex to think like a graph.
\newblock {\em Proceedings of the VLDB Endowment}, 7(3):193--204, 2013.

\bibitem{xie2014distributed}
C.~Xie, L.~Yan, W.-J. Li, and Z.~Zhang.
\newblock Distributed power-law graph computing: Theoretical and empirical
  analysis.
\newblock In {\em Advances in Neural Information Processing Systems}, pages
  1673--1681, 2014.

\bibitem{yan2014blogel}
D.~Yan, J.~Cheng, Y.~Lu, and W.~Ng.
\newblock Blogel: A block-centric framework for distributed computation on
  real-world graphs.
\newblock {\em Proceedings of the VLDB Endowment}, 7(14):1981--1992, 2014.

\bibitem{zhang2012accelerate}
Y.~Zhang, Q.~Gao, L.~Gao, and C.~Wang.
\newblock Accelerate large-scale iterative computation through asynchronous
  accumulative updates.
\newblock In {\em Proceedings of the 3rd workshop on Scientific Cloud
  Computing}, pages 13--22. ACM, 2012.

\end{thebibliography}

\end{document}